\documentclass[10pt,twocolumn,letterpaper]{article}

\usepackage{times}
\usepackage{epsfig}
\usepackage{graphicx}
\usepackage{amsmath}
\usepackage{amssymb}
\usepackage{algorithm}
\usepackage{algpseudocode}
\usepackage{relsize}
\usepackage{booktabs}

\usepackage{bbm}
\usepackage{multirow}
\usepackage{subcaption}
\usepackage[breaklinks=true,bookmarks=false]{hyperref}
\usepackage{natbib}

\newcommand{\argmin}[1]{\underset{#1}{\operatorname{arg}\,\operatorname{min}}\;}

\begin{document}
\setcounter{page}{1}
\title{Visualizing hierarchies in scRNA-seq data using a density tree-biased autoencoder}

\author{Quentin Garrido\,$^{\text{\sf 1,5}}$\thanks{\texttt{quentin.garrido[at]edu.esiee.fr}}\and
Sebastian Damrich\,$^{\text{\sf 1}}$\and
Alexander J{\"a}ger\,$^{\text{\sf 1}}$\and
Dario Cerletti\,$^{\text{\sf 2,3}}$ \and
Manfred Claassen\,$^{\text{\sf 4}}$ \and
Laurent Najman\,$^{\text{\sf 5}}$ \and
Fred A. Hamprecht\,$^{\text{\sf 1}}$\and\\
$^{\text{\sf 1}}$HCI/IWR, Heidelberg University, Germany\\
$^{\text{\sf 2}}$ Institute of Molecular Systems Biology, ETH Z{\"u}rich, Switzerland\\
$^{\text{\sf 3}}$ Institute of Microbiology, ETH Z{\"u}rich, Switzerland\\
$^{\text{\sf 4}}$ Internal Medicine I, University Hospital T{\"u}bingen, University of Tübingen, Germany\\
$^{\text{\sf 5}}$ Université Gustave Eiffel, CNRS, LIGM, F-77454 Marne-la-Vallée, France
}
\date{}
\maketitle

\begin{abstract}
\it
\textbf{Motivation:} Single cell RNA sequencing (scRNA-seq) allows studying the development of cells in unprecedented detail. Given that many cellular differentiation processes are hierarchical, their scRNA-seq data is expected to be approximately tree-shaped in gene expression space. Inference and representation of this tree-structure in two dimensions is highly desirable for biological interpretation and exploratory analysis.\\
\textbf{Results:}
Our two contributions are an approach for identifying a meaningful tree structure from high-dimensional scRNA-seq data, and a visualization method respecting the tree-structure. We extract the tree structure by means of a density based maximum spanning tree on a vector quantization of the data and show that it captures biological information well. We then introduce DTAE, a tree-biased autoencoder that emphasizes the tree structure of the data in low dimensional space. We compare to other dimension reduction methods and demonstrate the success of our method both qualitatively and quantitatively on real and toy data.\\
\textbf{Availability:} Our implementation relying on PyTorch \citep{paszke_pytorch_2019} and
Higra \citep{perret_higra_2019} is available at
\href{https://github.com/hci-unihd/DTAE}{https://github.com/hci-unihd/DTAE}.\\
\end{abstract}

\section{Introduction}

Single-cell RNA sequencing (scRNA-seq) data allows analyzing gene expression
profiles at the single-cell level, thus granting insights into cell behavior at unparalleled resolution. In particular, this permits studying the cell development through time more precisely.

Waddington's popular metaphor likens the development of cells to marbles rolling down a landscape~\citep{waddington_strategy_1957}. While cells are all grouped at the top of the hill when they are not yet differentiated (e.g., stem cells), as they start rolling down, they can take multiple paths and end up in distinct differentiated states, or cell fates.

However, for every cell, hundreds or thousands of expressed genes are recorded, and this data is noisy.
To summarize such high-dimensional data, it is useful to visualize it in two or three dimensions. 

\begin{figure*}[!tbp]
\centering
\includegraphics[width=\textwidth]{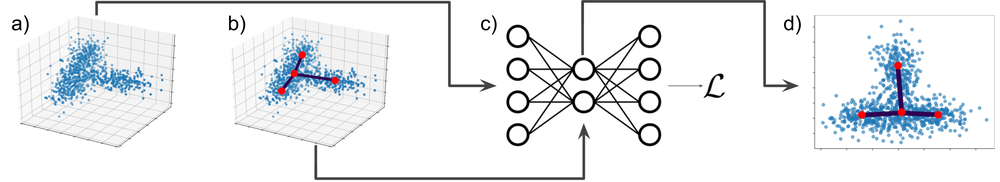}
\caption{Schematic method overview. \textbf{a}) High-dimensional data. \textbf{b}) Proposed density tree. After computing the $k$-means centroids on the data, we build a tree based on the data density between pairs of centroids. \textbf{c}) DTAE. An autoencoder is used to learn a representation of our data. This embedding is regularized by the previously computed tree in order to preserve its hierarchical structure in low-dimensional space. \textbf{d}) The final DTAE embedding. After training of the autoencoder, the bottleneck layer visualizes the data in low dimension and respects the density structure.} 
\label{fig:method}
\end{figure*}

Our goal, then, is to identify the hierarchical (tree) structure of the scRNA-seq data and subsequently reduce its dimensionality while preserving the extracted hierarchical properties. We address this in two steps, illustrated in figure~\ref{fig:method}.

First, we cluster the scRNA-seq data in high-dimensional space to obtain a more concise and robust representation. Then, we capture the hierarchical structure as a minimum spanning tree (MST) on our cluster centers, with edge weights reflecting the data density in high-dimensional space. We dub the resulting tree ``density tree''. 

Second, we embed the data to low dimension with an autoencoder, a type of artificial neural network. In addition to the usual aim of reconstructing its input, we bias the autoencoder to also reproduce the density tree in low-dimensional space. As a result, the hierarchical properties of the data are emphasized in our visualization.

\section{Related Work}
There are various methods for visualizing scRNA-seq data and trajectory inference, and many of them have been reviewed for instance in \cite{saelens_comparison_2019}. We therefore mention only some exemplary approaches here. 

\textbf{Graph only.} SCORPIUS \citep{cannoodt_scorpius_2016} was one of the first such methods. It is limited to linear topologies rather than trees. More versatile methods include SLINGSHOT \citep{street_slingshot_2018} and SPADE~\citep{bendall2011single, qiu2011extracting}. In contrast to our work, these three methods only provide a graph summary of the data, but not a 2D scatter plot. Similar to our density tree, SLINGSHOT and SPADE determine the hierarchical structure of the dataset as a MST on cluster centers. However, SLINGSHOT does not consider density. SPADE addresses the data density only by downsampling dense regions to equalize the data density. In particular, it does not inform the MST by the actual data density, which can be problematic, as illustrated in figure~\ref{fig:mst}. In contrast, we induce our density tree to have edges in high-density regions. 
PAGA \citep{wolf_paga_2019} produces primarily a graph summary of the data. It first clusters the $k$ nearest neighbor ($k$NN) graph of the data by modularity, and then places edges between clusters of high connectivity. Optionally, a layout of the PAGA graph can serve as initialization to other methods, such as UMAP. In our method, we connect clusters depending on the data density between two cluster centroids. Moreover, our proposed visualization is directly optimized to respect the density tree, while PAGA injects graph information only at the initialization of the visualization.

\textbf{Graph and visualization.} 
MONOCLE 2 \citep{qiu_reversed_2017} is more similar to our method, as it provides both a visualization and a hierarchical graph structure on a vector quantization of the data. The tree in MONOCLE~2 is inferred in conjunction with the embedding, while we learn it as a first step in high-dimensional space and consider the data density explicitly. As a result, our density tree depends only on the biological data but not the embedding initialization or dimension. MONOCLE~2 is conceptually promising, but empirically found to be often inferior to other methods, confer~\citep{moon_visualizing_2019}. Hence, we did not compare to MONOCLE 2. 

\textbf{Visualization only.} Most visualization methods do not provide a graph representation of the data. 
PHATE \citep{moon_visualizing_2019} is a recent approach which computes diffusion probabilities on the data before applying multidimensional scaling. 

The general purpose dimension reduction methods \mbox{t-SNE}~\citep{maaten_visualizing_2008}, UMAP~\citep{mcinnes_umap_2020, becht_dimensionality_2019} and ForceAtlas2~\citep{jacomy2014forceatlas2} are popular for visualizing scRNA-seq data. They aim to layout the $k$NN graph structure of the data with $t$-SNE focusing more on discrete clusters and ForceAtlas2 better representing the continuous structure~\citep{bohm2020attraction}. This often works well, but lacks the focus on hierarchies that our method provides. While the continuous focus of ForceAtlas2 seems apt to show differentiation processes in scRNA-seq datasets, we find that without a specific tree-prior the biologically interesting branching events are often poorly resolved.

Like DTAE, several recent methods for visualizing scRNA-seq data rely on neural networks. We describe them in the following. Many approaches are extensions of the autoencoder (AE) \citep{rumelhart_learning_1985}, a network which encodes the data to a lower dimensional latent space from which it tries to decode the input. A prominent member of this family is DCA \citep{eraslan2019single} which replaces the usual reconstruction loss by a count-based ZINB loss and aims at denoising scRNA-seq data. Its extension scDeepCluster \citep{tian2019clustering} jointly trains a clustering model in latent space. SAUCIE \citep{amodio_exploring_2019} is another popular AE method and addresses multiple tasks including batch effect removal and clustering, for which it uses a binary hidden layer. In order to exploit more relational information, scGAE \citep{luo2021scgae} uses a graph AE based on the $k$NN graph and achieves good visualization results both for clustered and continuous scRNA-seq data, but without our inductive prior of a hierarchical embedding or our explicit focus on data density. Topological autoencoders \citep{moor_topological_2020} are conceptually closest to our idea of retaining topological properties during dimension reduction. They compute the MST on all points, which produces less stable results than our density-based approach on cluster centroids.

Variational autoencoders (VAEs) \citep{kingma2013auto}, a generative AE version, have also been explored.
A popular VAE for scRNA-seq data is scVI \citep{lopez2018deep}, which explicitly models batch effects and library sizes.  Instead, scVAE \citep{gronbech2020scvae} investigates likelihood functions suitable for scRNA-seq data and proposes a clustering model in latent space. DR-A \citep{lin2020deep} apply adverserial training instead of the variational objective.
Finally, scvis is a VAE tailored to visualization \citep{ding2018interpretable} and uses a t-SNE-like regularization term in the latent space.

Ivis \citep{szubert2019structure} employs a triplet loss function and a siamese neural network instead of an AE to preserve the nearest neighbor relations in the visualization. 

Both scDeepCluster and scVAE shape the latent space into disconnected clusters, which is orthogonal to our goal of illustrating continuous developmental hierarchies. scVI, scGAE and scDeepCluster work with a latent space dimension larger than two and thus require an additional dimension reduction, typically with $t$-SNE, to visualize the data. 
 
Neither of the pure visualization methods aims to bring out the hierarchical properties often present in scRNA-seq dataset. In particular, they do not use the data density to infer lineages. None of them provide a graph summary of the data. Our contribution, however, is to supply the user with a tree-shaped graph summarizing the hierarchies along dense lineages in the data as well as a 2D embedding that respects this tree shape.

\section{Methods}

\subsection{Approximating the High-dimensional scRNA-seq Data with a Tree
}
To summarize the high-dimensional data in terms of a tree, the minimum spanning tree (MST) on the Euclidean distances is an obvious choice. This route is followed by~\cite{moor_topological_2020} 
who reproduce the MST obtained on their high-dimensional data in their low-dimensional embedding.
However, scRNA-seq data can be noisy, and a MST built on all of our data is very sensitive to noise. Therefore, we first run $k$-means clustering on the original
data, yielding more robust centroids for the MST construction and also reducing downstream complexity.

\begin{figure*}[!htbp]
\centering
\includegraphics[width=0.9\textwidth]{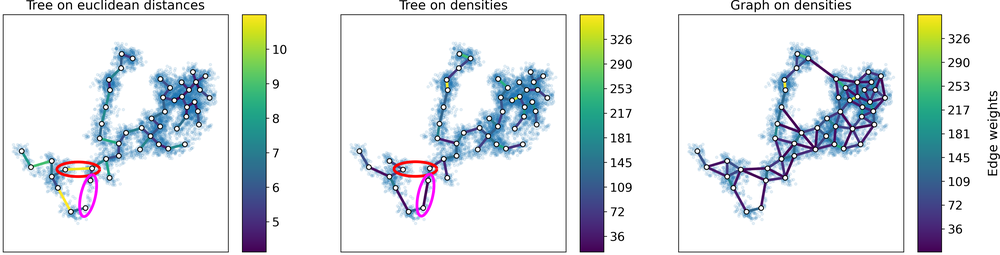}
\caption{(left, middle) Comparison of the tree built on $k$-means centroids using Euclidean distance or density weights. The data was generated using the PHATE library~\citep{moon_visualizing_2019}, with 3 branches in 2D. Original data points are transparently overlayed to better visualize their density. While the tree based on the Euclidean distance places
connections between centroids  that are close but have only few data points between them (see red ellipse), our tree based on the data density instead includes those edges that lie in high
density regions (see pink ellipse). (right) Complete graph over centroids and its Hebbian edge weights. Null-weight edges, that is edges not supported by data, are omitted for clarity.} 
\label{fig:mst}
\end{figure*}

	\begin{algorithm}[t]
		\begin{algorithmic}
		\caption{Density tree generation}
		\Require High-dimensional data $X \in \mathbb{R}^{n \times d}$
		\Require Number of $k$-means centroids $k$
		\Procedure{GenerateTree}{$X,k$}
			\State $C \gets \Call{kMeans}{X,k}$ \Comment{$\mathcal{O}(nkdt)$ with $t$ the number of
			iterations}
			\State $ G = (C,E)$ the complete graph on our centroids
			\For{$\{i,j\}$ a two-element subset of $\{1, \ldots, k\}$}  \Comment{$\mathcal{O}(k^2)$}
		        \State $d_{i,j} = 0$
			    \EndFor
			\For{$i = 1,\ldots, |X|$} \Comment{$\mathcal{O}(nk)$}
			    \State $a \gets \argmin{j=1, \ldots, k} ||x_i - c_j||_2$  \Comment{Nearest centroid}
			    \State $b \gets \argmin{\substack{j=1, \ldots, k \\  j\neq a}} ||x_i -c_j||_2 $ \Comment{Second nearest centroid}
			    \State $d_{a,b} = d_{a,b} + 1$ \Comment{Increase nearest centroids' edge strength}
			\EndFor
			\State $T \gets \Call{MaxSpanningTree}{G, d}$ \Comment{$\mathcal{O}(k^2\log{k})$}
			\State \Return $T, d$ \Comment{Retains the density tree and the edge strengths}
		\EndProcedure
		\label{algo:tree}
		\end{algorithmic}
	\end{algorithm}
A problem with the Euclidean MST, illustrated in figure~\ref{fig:mst}, is that two centroids can be close in Euclidean space without having many data points between them. In such a case, a Euclidean MST would not capture the skeleton of our original data well.
But it is crucial that the extracted tree follows the dense regions of the data if we want to visualize developmental trajectories of differentiating cells: a trajectory is plausible if we observe intermediate cell states and unlikely if there are jumps in the development.
By preferring tree edges in high density regions of the data, we ensure that the
computed spanning tree is biologically plausible.
Following this rationale, we build the maximum spanning tree on the complete graph over centroids whose edge weights are given by the density of the data along each edge instead of the minimum spanning tree on Euclidean distance. This results in a tree that (we believe) captures Waddington's hypothesis better than merely considering cumulative differences in expression levels.

To estimate the support that a data sample provides for an edge, we follow~\cite{martinetz_topology_1994}. Consider the complete graph $G = (C,E)$ such that $C = \{c_1 , \ldots, c_k\}$ is the set of centroids. In the spirit of Hebbian learning, that is, emphasizing connections that appear frequently, we count, for each edge, how often its incident vertices are the two closest centroids to any given datum. 

As pointed out by~\cite{martinetz_topology_1994} this amounts to an empirical estimate of the integral of the density of observations across the second-order Vorono\"i region
(defined as the set of points having a particular set of 2 centroids as its 2 nearest centroids)
associated with this pair of cluster centers. Finally, we compute the maximum spanning tree over these Hebbian edge weights.
Our strategy for building the tree is summarized in algorithm~\ref{algo:tree}.

Our data-density based tree follows the true shape of the data more closely than a MST based on the
Euclidean distance weights, as illustrated in figure~\ref{fig:mst}. We claim this indicates it being a better choice for capturing developmental trajectories.
Having extracted the tree shape in high dimensions, our goal is to reproduce this tree as closely as possible in our embedding.

\subsection{Density-Tree biased Autoencoder (DTAE)}

We use an autoencoder to faithfully embed the high-dimensional scRNA-seq data in a low-dimensional space, and bias it such that the topology inferred in high-dimensional space is respected. An autoencoder is an artificial neural network consisting of two concatenated subnetworks, the encoder $f$, which maps the input to lower-dimensional space, also called embedding space, and the decoder $g$, which tries to reconstruct the input from the lower-dimensional embedding. It can be seen as a non-linear generalization of PCA. We visualize the low-dimensional embeddings  $h_i=f(x_i)$ and hence choose their dimension to be $2$. 

The autoencoder is trained by minimizing the following loss terms, including new ones that bias the autoencoder to also adhere to the tree structure.

\subsubsection{Reconstruction Loss}
The first term of the loss is the reconstruction loss, defined as

\begin{equation}
	\mathcal{L}_{\text{rec}} = \text{MSE}(X, g(f(X))) = \frac{1}{N} \sum_{x_i \in X} ||x_i-g(f(x_i))||_2^2.
\end{equation}
This term is the typical loss function for an autoencoder and ensures that the embedding
is as faithful to the original data as possible, forcing it to extract the
most salient data features.

\subsubsection{Push-Pull Loss}
The main loss term that biases the DTAE towards the density tree is the push-pull
loss. It trains the encoder to embed the data points such that the high-dimensional data density, and, in particular, the density tree, are reproduced in low-dimensional-space.

We find a centroid in embedding space by averaging the embeddings of all points assigned to the corresponding \mbox{$k$-means} cluster in high-dimensional space. In this way, we can easily relate the centroids in high and low dimension, and will simply speak of centroids when the ambient space is clear from the context.

To reproduce the density structure in low-dimensional space, we want that the closest two high-dimensional centroids to a point $x_i \in X$ correspond to the two low-dimensional centroids that are closest to its embedding \mbox{$h_i = f(x_i)$}. We denote the latter centroids by $c_{i,1}$ and $c_{i,2}$, and low-dimensional centroids that actually correspond to the closest high-dimensional centroids by $c'_{i,1}$ and $c'_{i,2}$. 
As long as $c'_{i,1}$, $c'_{i,2}$ differ from $c_{i,1}$ and $c_{i,2}$, the encoder places $h_i$ next to different centroids than in high-dimensional space. To improve this, we want to move $c'_{i,1}$, $c'_{i,2}$ and $h_i$ towards each other while separating $c_{i,1}$ and $c_{i,2}$ from $h_i$. The following preliminary version of our push-pull loss implements this:

\begin{align}
    \tilde{\mathcal{L}}_{\text{push}}(h_i)&= -\left(||h_i - c_{i,1}||_2 + ||h_i - c_{i,2}||_2\right)^2  \\
    \tilde{\mathcal{L}}_{\text{pull}}(h_i)&= \left(||h_i - c'_{i,1}||_2 + ||h_i - c'_{i,2}||_2\right)^2 \\
	\tilde{\mathcal{L}}_{\text{push-pull}} &= \frac{1}{N} \sum_{x_i \in X}
	\tilde{\mathcal{L}}_{\text{push}}(f(x_i)) + \tilde{\mathcal{L}}_{\text{pull}}(f(x_i)).\label{eq:first_ppLoss}
\end{align}
The push loss decreases as $h_i$ and the currently closest centroids, $c_{i,1}$ and $c_{i,2}$, are placed further apart from each other, while the pull loss decreases when $h_i$ gets closer to the correct centroids, $c'_{i,1}$ and $c'_{i,2}$. Indeed, the push-pull loss term is minimized if and only if each embedding $h_i$ lies in the second-order Vorono\"i region of those low-dimensional centroids whose high-dimensional counterparts contain the data point $x_i$ in their second-order Vorono\"i region. In other words, the loss is zero precisely when we are reproducing the edge densities from high dimension in low dimension.

Note that we let the gradient flow through both the individual embeddings and through the centroids, which are means of embeddings themselves.

This na\"ive formulation of the push-pull loss has the drawback that it can become very small if all embeddings are nearly collapsed into a single point, which is undesirable for visualization. Therefore, we normalize the contribution of every embedding $h_i$ by the distance between the
two correct centroids in embedding space. This prevents the collapsing of embeddings, and also ensures that each datapoint $x_i$ contributes equally, regardless of how far apart
$c'_{i,1}$ and $c'_{i,2}$ are. The push-pull loss thus becomes
\begin{align}
    \mathcal{L}_{\text{push}}(h_i) &= -\left(\frac{||h_i - c_{i,1}||_2 +
			||h_i - c_{i,2}||_2}{||c'_{i,1} - c'_{i,2}||_2}\right)^2  \\
    \mathcal{L}_{\text{pull}}(h_i) &= \left(\frac{||h_i - c'_{i,1}||_2 + 
	||h_i - c'_{i,2}||_2}{||c'_{i,1} - c'_{i,2}||_2}\right)^2 \\
	\mathcal{L}_{\text{push-pull}} &= \frac{1}{N}\sum_{x_i \in X}
	\mathcal{L}_{\text{push}}(f(x_i)) + \mathcal{L}_{\text{pull}}(f(x_i)). \label{eq:unweighted_pploss}
\end{align}

So far, we only used the density information from high-dimensional space for the embedding, but not the extracted density tree itself. The push-pull loss in equation \eqref{eq:unweighted_pploss} is agnostic to the positions of the involved centroids within the density tree, only their Euclidean distance to the embedding $h_i$ matters. In contrast, the hierarchical structure is important for the biological interpretation of the data: it is much less important if an embedding is placed close to two centroids that are on the same branch of the density tree than it is if the embedding is placed between two different branches. In the first case, cells are just not ordered correctly within a trajectory, while in the second case we get false evidence for an altogether different pathway. 

We tackle this problem by reweighing the push-pull loss with the geodesic distance along the density tree. The geodesic distance $d_{\text{geo}}(c_a,c_b)$ with $c_a,c_b \in C$ is defined as the number of edges in the shortest path between $c_a$ and $c_b$ in the density tree. 
Centroids at the end of different branches in the density have a higher geodesic distance than centroids nearby on the same branch. By weighing the push-pull loss contribution of an embedded point by the geodesic distance between its two currently closest centroids, we focus the push-pull loss on embeddings which erroneously lie between different branches.

The geodesic distances can be computed quickly in $\mathcal{O}(k^2)$ via breadth first search, and this only has to be done once before training the autoencoder.

The final version of our push-pull loss becomes
\begin{align}
	\mathcal{L}_{\text{push-pull}} = \frac{1}{N} \sum_{x_i \in X} \Big(
	&d_{\text{geo}}(c_{i,1},c_{i,2})\nonumber \\ 
	&\cdot \left(\mathcal{L}_{\text{push}}(f(x_i))+
	\mathcal{L}_{\text{pull}}(f(x_i))\right)\Big). \label{eq:final_pploss}
\end{align}
Note, that the normalized push-pull loss in equation \eqref{eq:unweighted_pploss} and the geodesically reweighted push-pull loss in \eqref{eq:final_pploss} both also get minimized if and only if the closest centroids in embedding space  correspond to the closest centroids in high-dimensional space.

\subsubsection{Compactness loss}
The push-pull loss replicates the empirical high-dimensional data density in embedding space by moving the embeddings into the correct second-order Vorono\"i region, which can be large or unbounded. For optimal visibility of the tree structure, an embedding should not only be in the correct second-order Vorono\"i region, but lie compactly around the line between its two centroids. To achieve this, we add the compactness loss, which is just another instance of the pull loss
\begin{align}
	\mathcal{L}_{\text{comp}}&= \frac{1}{N} \sum_{x_i \in
	X}\left(\frac{||h_i - c'_{i,1}||_2 + ||h_i - c'_{i,2}||_2}{||c'_{i,1} - c'_{i,2}||_2}\right)^2\\
	&= \frac{1}{N} \sum_{x_i \in X} \mathcal{L}_{\text{pull}}(f(x_i)),
\end{align}
The compactness loss is minimized if the embedding $h_i$ is exactly between the correct centroids $c'_{i,1}$ and $c'_{i,2}$ and has elliptic contour lines with foci at the centroids.

\subsubsection{Cosine loss}
Since the encoder is a powerful non-linear map, it can introduce artifactual curves in the low-dimensional tree branches. However, especially tight turns can impede the visual clarity of the embedding. As a remedy, we propose an optional additional loss term that tends to straighten branches.

Centroids at which the embedding should be straight are the ones within a branch, but not at a branching event of the density tree. The former can easily be identified as the centroids of degree 2.

Let $c$ be a centroid in embedding space of degree 2 with its two neighboring centroids $n_{c,1}$ and
$n_{c,2}$. The branch is straight at $c$ if the two vectors $c - n_{c,1}$ and $n_{c,2} - c$ are parallel or, equivalently, if their cosine is maximal. Denoting by $C_2 = \{c \in C\;|\; \text{deg}(c) = 2\}$ the set of all centroids of degree 2, considered in embedding space, we define the cosine loss as 
\begin{equation}
	\mathcal{L}_{\text{cosine}} = 1 - \frac{1}{|C_2|} \mathlarger{\sum_{c \in C_2}} \frac{(c-n_{c,1})
		\cdot (n_{c,2}-c)}{||c-n_{c,1}||_2\;
	||n_{c,2}-c||_2}. 
\end{equation}
Essentially, it measures the cosine of the angles along the tree branch and becomes minimal if all these angles are zero and the branches straight.

A generalization of this criterion that deals with noisy edges in the density tree is discussed in section B of the appendix.

\subsubsection{Complete loss function}
Combining the four loss terms of the preceding sections, we arrive at our final loss
\begin{equation}
	\mathcal{L} = \lambda_{\text{rec}}\mathcal{L}_{\text{rec}} + \lambda_{\text{push-pull}}\mathcal{L}_{\text{push-pull}} +
	\lambda_{\text{comp}}\mathcal{L}_{\text{comp}} + \lambda_{\text{cos}}\mathcal{L}_{\text{cos}}.
\end{equation}
The relative importance of the loss terms, especially of $\mathcal{L}_{\text{comp}}$ and $\mathcal{L}_{\text{cos}}$, which control finer aspects of the visualization, might depend on the use-case. In practice, we found $\lambda_{\text{rec}} = \lambda_{\text{push-pull}} =\lambda_{\text{comp}} = 1$ and
$\lambda_{\text{cos}} = 50$ to work well. This configuration reduces the number of weights to adjust from four to one.

An ablation study of the different losses' contribution is available in section C of the appendix. Its main conclusion is that while the push-pull loss and reconstruction loss are sufficient to obtain satisfactory results, the addition of the compactness and cosine loss helps to improve the visualizations further and facilitates reproducibility. Empirically, we found that adding the compactness loss without the cosine loss sometimes leads to discontinuous embeddings. The two loss terms should therefore be added or omitted jointly.

If the default loss weights are not satisfactory, we recommend adjusting the cosine loss weight first. To understand how changing the loss parameters may affect the results, please refer to the qualitative results in the ablation study.

\subsection{Training procedure}

\begin{figure*}[t]
	\centering
	\includegraphics[width=\linewidth]{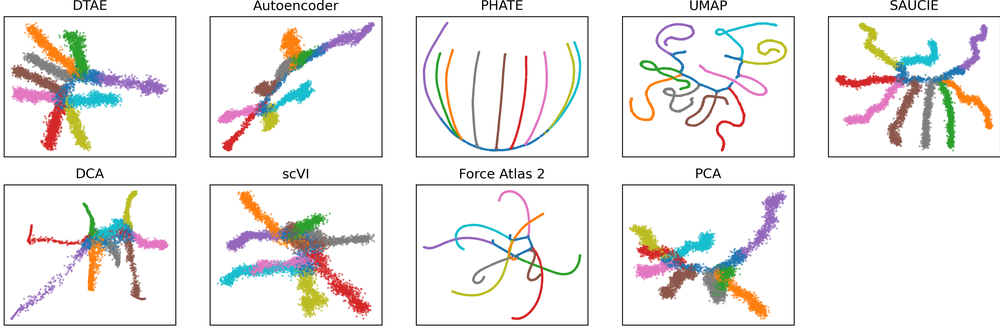}
	\caption{Results obtained using data generated by the PHATE library. Branches are coloured by groundtruth labels.}
	\label{fig:results_phate}
\end{figure*}

Firstly, we compute the $k$-means centroids, the edge densities, the density tree, and geodesic distances. This has to be done only once as an initialization step. Secondly, we pretrain the autoencoder with only the reconstruction loss via stochastic gradient descent on minibatches. This provides a warm start for finetuning the autoencoder with all losses in the third step.

During finetuning, all embedding points are needed to compute the centroids in embedding space. Therefore, we perform full-batch gradient descent during finetuning. For algorithmic details regarding the training procedure, confer to supplementary algorithm~\ref{algo:training}.

We always used $k=50$ centroids for $k$-means clustering in our experiments. This number needs to be high enough so that the tree yields a skeleton of the data, but not so high that the density loses its meaning. $k=50$ is a default value that works well in a variety of scenarios.
Our autoencoder always has a bottleneck dimension of 2 for visualization. In the experiments, we used layers of the following dimensions $d (\textnormal{input dimension}), 2048, 256,32,2,32,256,2048, d$. This results in symmetrical encoders and decoders with four layers. While not necessary in our experiments, if a lighter network is desired, we recommend applying PCA first to reduce the number of input dimensions, or to filter out more genes during the preprocessing.
We omitted hidden layers of dimension larger than the input.  We use fully connected layers and ReLU activations after every layer but the last encoder and decoder layer and employ the Adam~\citep{kingma_adam_2017} optimizer
with learning rate $2\times10^{-4} $ for pretraining and $1\times10^{-3} $ for finetuning unless stated otherwise. We used a batch size of $256$ for pretraining in all experiments.

\section{Results}
In this section, we show the performance of our method on toy and real scRNA-seq datasets and compare it to a vanilla autoencoder, as well as to the popular non-parametric methods PCA, Force Atlas 2, UMAP and PHATE and to the most prevalent neural network-based approaches, SAUCIE, DCA and scVI. For all network-based approaches, we choose a bottleneck of dimension 2 to directly use them for visualization.

\subsection{PHATE generated data}

We applied our method to an artificial dataset created with the library published alongside~\cite{moon_visualizing_2019}, to demonstrate its functionality in a controlled setting. We generated a toy dataset whose skeleton is a tree with one backbone branch and 9 branches emanating from the backbone, consisting in total of 10,000 points in 100 dimensions. 

We pretrained for 150 epochs with a learning rate of $10^{-3}$ and finetuned for another 150 epochs with a learning rate of $10^{-2}$.

Figure~\ref{fig:results_phate} shows the visualization results. The finetuning significantly improves the results of the pretrained autoencoder, whose visualisation collapses the grey and green branch onto the blue branch. All methods other than DCA, scVI and PCA achieve satisfactory results that make the true tree structure of the data evident. While PHATE, UMAP and Force Atlas 2 produce overly crisp branches compared to the PCA result, the reconstruction loss of our autoencoder guards us from collapsing the branches into lines. PHATE appears to overlap the cyan and yellow branches near the backbone, and UMAP introduces artificially curved branches. scVI collapses the green and brown as well as the pink and cyan branches together, giving hard to interpret visualizations.
The results on this toy dataset demonstrate that our method can embed high-dimensional hierarchical data into 2D and emphasize its tree-structure while avoiding to collapse too much information compared to state-of-the-art methods. In our method, all branches are easily visible.
\begin{figure*}[t]
	\centering 
	\includegraphics[width=1\linewidth]{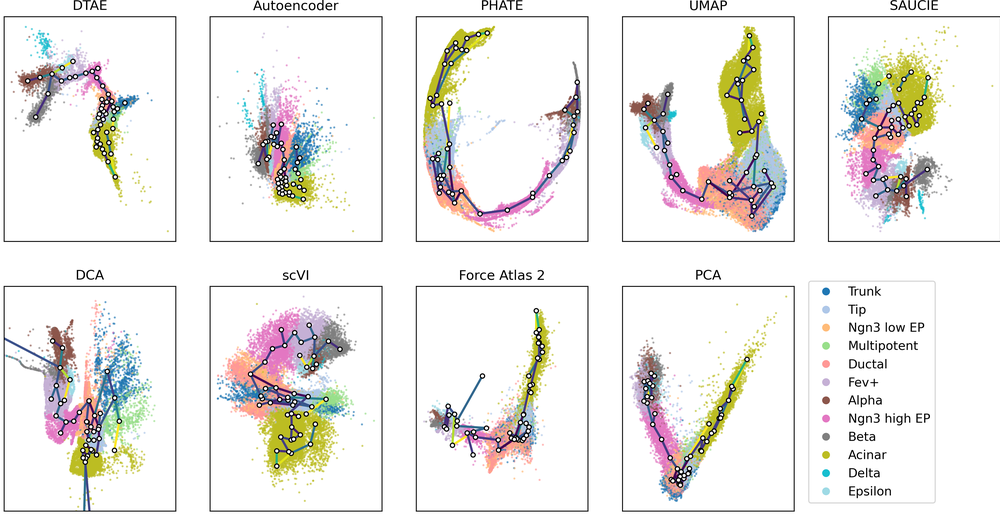}
	\caption{Pruned density tree superimposed over embeddings of the endocrine pancreatic cell dataset, colored by cell subtypes. We use finer labels for the endocrine cells. Darker edges represent denser edges.
	Only edges with more than 100 points contributing to them are plotted here.}%
	\label{fig:results-pan-tree} 
\end{figure*}

\subsection{Endocrine pancreatic cell data}

We evaluated our method on the data from~\cite{bastidas-ponce_comprehensive_2019}. It represents endocrine pancreatic cells at different stages of their development and consists of gene expression information for 36351 cells and 3999 genes. Preprocessing information can be found in~\cite{bastidas-ponce_comprehensive_2019}. We pretrained for 300 epochs and used 250 epochs for finetuning.

Figure~\ref{fig:results-pan-tree} and supplementary figure~\ref{fig:results-pancreas} depicts visualizations of the embryonic pancreas
development with different methods. Our method can faithfully reproduce the tree structure of the data, especially for the endocrine subtypes. The visualized hierarchy is biologically plausible, with a particularly clear depiction of the $\alpha$-, $\beta$- and $\varepsilon$-cell branches and a visible, albeit too strong, separation of the $\delta$-cells. This is in agreement with the results from~\cite{bastidas-ponce_comprehensive_2019}. UMAP also performs very well and attaches the $\delta$-cells to the main trajectory. However, the $\alpha$- and $\beta$-cell branches are not as prominent as in DTAE. PHATE does not manage to separate the $\delta$- and $\varepsilon$-cells discernibly from the other endocrine subtypes. As on toy data in figure~\ref{fig:results_phate}, it produces overly crisp branches for the $\alpha$- and $\beta$-cells. PCA mostly overlays all endocrine subtypes. All methods but the vanilla autoencoder show a clear branch with tip and acinar cells and one via EP and Fev+ cells to the endocrine subtypes, but only DTAE, DCA, SAUCIE and scVI manage to also hint at the more generic trunk and multipotent cells from which these two major branches emanate. 
However, SAUCIE, DCA and scVI fail to produce a meaningful separation between the $\alpha$- and $\beta$-cell branches.
The ductal and Ngn3 low EP cells overlap in all methods.

It is worth noting that the autoencoder alone was not able to visualize meaningful hierarchical properties of the data. However, the density tree-biased finetuning in DTAE made this structure evident, highlighting the benefits of our approach.

In figure~\ref{fig:results-pan-tree}, we overlay DTAE's embedding with a pruned version of the density tree and see that the visualization closely follows the tree structure around the differentiated endocrine cells. This combined representation of low-dimensional embedding and overlaid density tree further facilitates the identification of branching events, most notably for the $\alpha$- and $\beta$-cells, and shows the full power of our method. It also provides an explanation for the apparent separation of the $\delta$-cells. Since there are relatively few $\delta$-cells, they are not represented by a distinct $k$-means centroid.

\begin{figure*}[t]
	\centering 
	\includegraphics[width=1\linewidth]{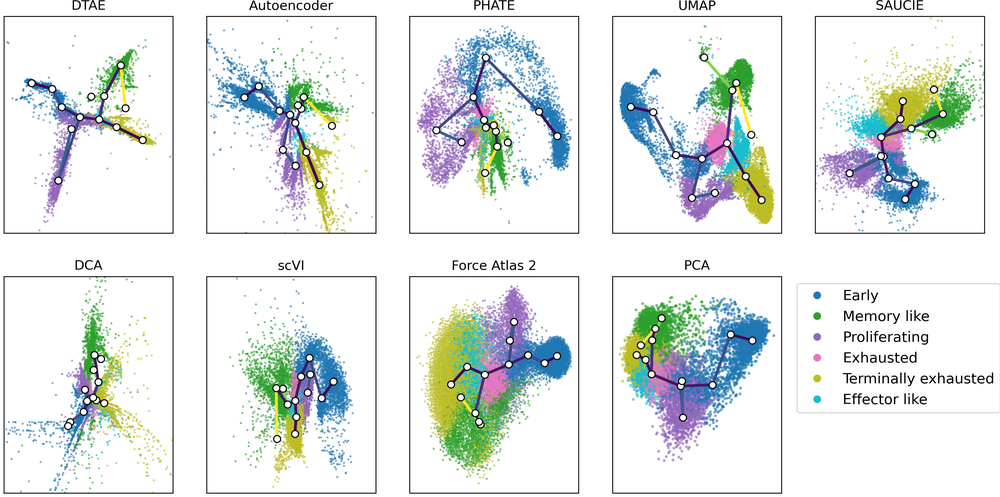}
	\caption{Pruned density tree superimposed over embeddings of the chronic part of the T-cell 
	data, colored by phenotypes. Darker edges represent denser edges.
	Only edges with more than 100 points contributing to them are plotted here.}%
	\label{fig:results-eth-tree} 
\end{figure*}

Our method places more $k$-means centroids in the dense region in the lower right part of DTAE's panel in figure~\ref{fig:results-pan-tree} than is appropriate to capture the trajectories, resulting in many small branches. Fortunately, this does not result in an exaggerated tree-shaped visualization that follows every spurious branch, which we hypothesize is thanks to the successful interplay between the tree bias and the reconstruction aim of the autoencoder: If the biological signal encoded in the gene expressions can be reconstructed by the decoder from an embedding with enhanced hierarchical structure, the tree-bias shapes the visualization accordingly. Conversely, an inappropriate tree-shape is prevented if it would impair the reconstruction. Overall, the density tree recovers the pathways identified in~\cite{bastidas-ponce_comprehensive_2019} to a large extent. Only the trajectory from multipotent via tip to acinar cells includes an unexpected detour via the trunk and ductal cells, which the autoencoder mends by placing the tip next to the multipotent cells. 

The density tree also provides useful information in conjunction with other dimension reduction methods. In figure~\ref{fig:results-pan-tree}, we overlay their visualizations with the pruned density tree by computing the centroids in the respective embedding spaces according to the $k$-means cluster assignments. The density tree can help to find branching events and gain insights into the hierarchical structure of the data that is visualized with an existing dimension reduction method. For instance, together with the density tree, we can identify the $\varepsilon$-cells as a separate branch and find the location of the branching event into different endocrine subtypes in the UMAP embedding.

\subsection{T-cell infection data}

We further applied our method to T-cell data of a chronic and an acute infection, which was shared with us by the authors of~\cite{cerletti_fate_2020}. The data was preprocessed using the method described in~\cite{zheng_massively_2017}, for more details confer~\cite{cerletti_fate_2020}. It contains gene expression information for 19029 cells and 4999 genes. While we used the combined dataset to fit all dimension reduction methods, we only visualize the 13707 cells of the chronic infection for which we have phenotype annotations from~\cite{cerletti_fate_2020} allowing us to judge visualization quality from a biological viewpoint. We pretrained for 600 epochs and used 250 epochs for finetuning.

Figure~\ref{fig:results-eth-tree} and supplementary figure~\ref{fig:results-eth-chronic-louvain} demonstrate that our method makes the tree structure of the data clearly visible. The visualized hierarchy is also biologically significant: The two branches on the right correspond to the memory-like and terminally exhausted phenotypic states, which are identified as the main terminal fates of the differentiation process in~\cite{cerletti_fate_2020}. Furthermore, the purple branch at the bottom contains the proliferating cells. Since the cell cycle affects cell transcription significantly, those cells are expected to be distinct from the rest.

\begin{table*}[t!]
\centering
         \begin{tabular}{l c c c c c c c c c} 
         \toprule
          Type of metric & \multicolumn{2}{c}{Local} & \multicolumn{4}{c}{Global} & \multicolumn{2}{c}{Voronoi} & \\
           \cmidrule(l{2pt}r{2pt}){2-3} \cmidrule(l{2pt}r{2pt}){4-7} \cmidrule(l{2pt}r{2pt}){8-9} 
          \multirow{2}{*}{Metric}&\multirow{2}{*}{ARI}&\multirow{2}{*}{k-NN}& \multicolumn{2}{c}{Euclidean} & \multicolumn{2}{c}{Geodesic} &\multirow{2}{*}{1\textsuperscript{st} order} & \multirow{2}{*}{2\textsuperscript{nd} order} & All \\
         \cmidrule(l{2pt}r{2pt}){4-5} \cmidrule(l{2pt}r{2pt}){6-7} 
           &  & & Pearson & Spearman & Pearson & Spearman &   &  \\
         \hline
        DTAE (Ours) & \textbf{93.75}  & 48.70  & 85.51  & 72.91  & 82.39  & 87.19  & \textbf{98.24}  & \textbf{94.21}  & \textbf{82.86} \\ 
AE & 74.83  & 70.96  & 87.41  & 77.20  & 70.16  & 73.23  & 89.83  & 58.43  & 75.26 \\ 
PHATE & 84.76  & 73.48  & 45.43  & 46.04  & 74.15  & 78.45  & 85.27  & 44.04  & 66.45 \\ 
UMAP & 78.88  & \textbf{87.75}  & 53.42  & 54.31  & 79.40  & 80.12  & 83.31  & 55.94  & 71.64 \\ 
SAUCIE & 89.99  & 67.43  & 82.22  & 78.50  & 84.03  & 85.41  & 96.43  & 78.58  & 82.83 \\ 
DCA & 49.79  & 64.37  & 76.54  & \textbf{90.95}  & 40.40  & 65.92  & 63.26  & 49.33  & 62.57 \\ 
scVI & 74.80  & 54.30  & \textbf{87.82}  & 67.68  & 75.45  & 82.75  & 86.42  & 57.77  & 73.37 \\ 
Force Atlas 2 & 72.88  & 72.23  & 37.28  & 48.06  & 35.67  & 76.65  & 77.27  & 43.27  & 57.91 \\ 
PCA & 60.40  & 40.78  & 73.42  & 66.02  & \textbf{96.44}  & \textbf{96.40}  & 80.76  & 56.82  & 71.38 \\ 

 \bottomrule
 \end{tabular}
 \caption{Relative quantitative performances averaged over all studied datasets. For each metric, we give the best performing method a value of 100 and scale other results proportionally. The metrics are described in section~\ref{sec:quantitative} and higher values indicate better performance. The rightmost column contains the average relative performance over all metrics. DTAE and SAUCIE have the best performance overall, with DTAE excelling in Voronoï metrics and ARI.}
 \label{tab:quantitative-aggregated}
\end{table*}

It is encouraging that DTAE makes the expected biological structure apparent even without relying on known marker genes or differential cell expression, which were used to obtain the phenotypic annotations in~\cite{cerletti_fate_2020}.

Interestingly, our method places the branching event towards the memory-like cells in the vicinity of the exhausted cells, as does UMAP, while~\cite{cerletti_fate_2020} recognized a trajectory directly from the early stage cells to the memory-like fate. The exact location of a branching event in a cell differentiation process is difficult to determine precisely. We conjecture that fitting the dimensionality reduction methods on the gene expression measurements of cells from an acute infection in addition to those from the chronic infection analyzed in~\cite{cerletti_fate_2020} provided additional evidence for the trajectory via exhausted cells to the memory-like fate. Unfortunately, an in-depth investigation of this phenomenon is beyond the scope of this methodological paper.

The competing methods expose the tree-structure of the data less obviously than DTAE. The finetuning significantly improves the results from the autoencoder, which shows no discernible hierarchical structure. PHATE separates the early cells, proliferating cells and the rest. But its layout is very tight around the biologically interesting branching event towards memory-like and terminally exhausted cells. PCA exhibits only the coarsest structure and fails to separate the later states visibly. The biological structure is decently preserved in the UMAP visualization, but the hierarchy is less apparent than in DTAE. SAUCIE, scVI and Force Atlas 2 produce results that are very similar to PCA, with later states that are hard to distinguish. DCA produces results that are very similar to the vanilla autoencoder, where even though the later states are visible, there is a significant amount of noise in the embedding, making the analysis difficult.
Overall, our method outperforms the other visualization methods on this dataset.

In figure~\ref{fig:results-eth-tree}, we have overlaid our embedding with a pruned version of the density tree and see that DTAE's visualization indeed closely follows the tree structure. It is noteworthy that even the circular behavior of proliferation cells is accurately captured by a self-overlaid branch, although our tree-based method is not directly designed to extract circular structure.

Figure~\ref{fig:results-eth-tree} also shows the other dimension reduction methods in conjunction with the pruned density tree. Reassuringly, we find that all methods embed the tree in a plausible way, i.e., without many self-intersections or oscillating branches. This is evidence that our density tree indeed captures a meaningful tree structure of the data. As for the endocrine pancreas dataset, the density tree can enhance hierarchical structure in visualizations of existing dimension reduction methods. It, for example, clarifies in the UMAP plot that the pathway towards the terminally exhausted cells is via the exhausted and effector like cells and not directly via the proliferating cells.

\subsection{Quantitative analysis}\label{sec:quantitative}

The purpose of a visualization method is to make the most salient, qualitative properties of a dataset visible. Nevertheless, a quantitative evaluation can support the comparison of visualization methods and provide evidence that the data and its visualization are structurally similar. Unfortunately, there is to our knowledge no consensus as to which metric aligns with practitioners' notion of a useful visualization. Hence, any single metric cannot validate the quality of a method.
This is why it is important to use multiple metrics, so that one can hope for a more reliable result. 

We selected eight different metrics, some of which have been employed to judge visualization methods before~\citep{moon_visualizing_2019, kobak_art_2019,becht_dimensionality_2019}. The first group of metric considers the local structure. We compute the Adjusted Rand Index (ARI) between a $k$-means clustering in high and low dimension and the number of correct neighbors in the $k$-NN graph in high and low dimension.
The next category are global metrics, which rely on distance preservation. Euclidean distances are computed in low dimension and euclidean or geodesic distances are computed in high dimension. Then correlations are computed between those distances. 
Finally, we use Voronoï diagram based metrics. First or second order Voronoï diagrams on the $k$-means centroids are computed using the $k$-means assignments to obtain the seeds in low-dimensional space. Then the ratio of points placed in the correct Voronoï region is computed. When using the second order Voronoï diagram with $k=50$, there is a bias towards DTAE since we optimize this criterion.
For local and Voronoï diagram based metrics, we have to adjust a parameter $k$ (either for $k$-means clustering or for a $k$-NN graph). We vary the value of $k$ between 10 and 100 with a step of 10 and report the area under the curve.

We report results aggregated on all three datasets in table~\ref{tab:quantitative-aggregated} and full results are available in supplementary table~\ref{tab:quantitative-full}. This aggregation makes it easier to deduce general patterns of performance among multiple datasets. From the results on all datasets, we can clearly see that DTAE outperforms other methods on Voronoï diagram based metrics, in part due to the bias towards them for $k=50$. 
On local metrics, DTAE achieves the best performance on ARI, followed closely by SAUCIE. However, for $k$-NN preservation UMAP performs better than other methods by a significant margin which is consistent with the criterion it optimizes~\citep{damrich_umaps_2021}. %
For euclidean distance preservation, autoencoder based methods perform the best, with no clear winner overall.
For geodesic distance preservation, PCA performs the best, even though it produced poor visualizations. This is in line with previous findings~\citep{kobak_art_2019}. Most other methods obtained very similar performance on this metric, making it hard to conclude that any method performs better than another.

In order to more easily compare methods, aggregated performances over all metrics are reported in the rightmost column of table~\ref{tab:quantitative-aggregated}. This aggregation makes it easier to evaluate the overall performance of a method when using a wide variety of criteria. We chose the arithmetic mean to combine the results for simplicity's sake. 
From this, we can see that DTAE and SAUCIE perform significantly better than other methods, with DTAE surpassing SAUCIE by a small margin. However, from a qualitative point of view, DTAE produced superior visualizations compared to SAUCIE, as discussed previously.

Overall, DTAE produced excellent results both from a quantitative and qualitative point of view, highlighting its usefulness as a visualization method for tree-shaped data.

\section{Limitations}

\subsection{Hierarchy assumption}
\begin{figure}[t!]
	\centering  
	\includegraphics[width=1\linewidth]{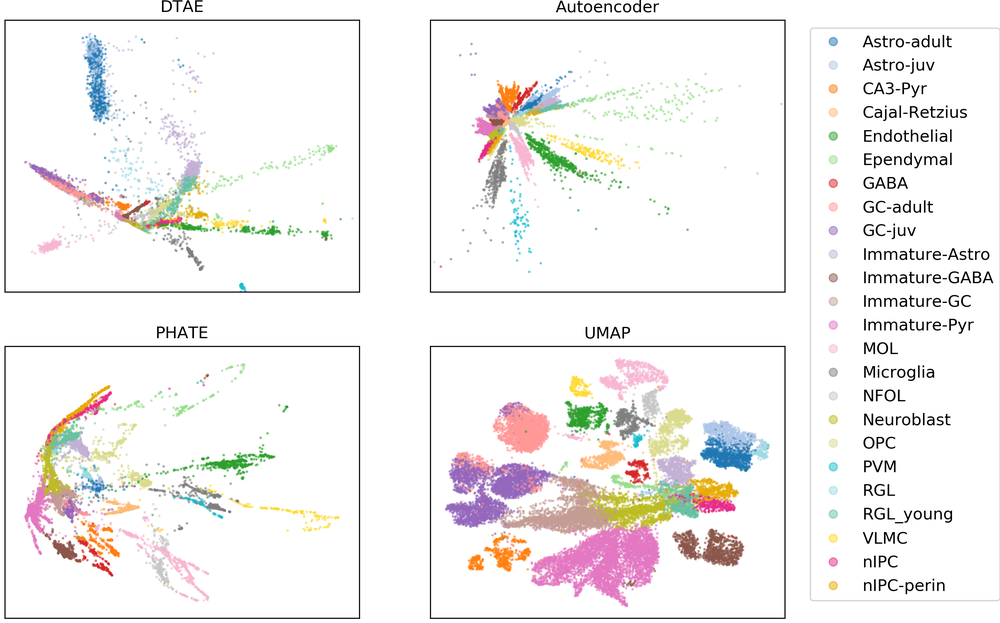}
	\caption{Failure case: Highly clustered data violates our underlying assumption of a tree structure. Dentate gyrus data from~\cite{hochgerner_conserved_2018} with clusters colored by groundtruth cluster assignments.}
	\label{fig:results_dentate} 
\end{figure}

Our method is tailored to Waddington's hierarchical structure assumption of developmental cell populations, in which the highest data density is along the developmental trajectory. It produces convincing results in this setting as shown above. However, if the assumption is violated, for instance because the dataset contains multiple separate developmental hierarchies or a mixture of hierarchies and distinct clusters of fully differentiated cell fates, the density tree cannot possibly be a faithful representation of the dataset. Indeed, in such a case, our method yields a poor result. As an example, confer figure~\ref{fig:results_dentate} with visualizations of the dentate gyrus dataset from~\cite{hochgerner_conserved_2018}, preprocessed according to~\cite{zheng_massively_2017}. This dataset consists of a mostly linear cell trajectory and several distinct clusters of differentiated cells, and consequently does not meet our model's assumption. Indeed, DTAE manages to only extract some linear structures, but overall fails on this dataset, similarly to PHATE. UMAP seems to produce the most useful visualization here.

One could adapt our method by extracting a forest of disconnected density trees by cutting edges below a density threshold. However, if little is known a priori about the structure of the dataset, a more general dimension reduction method might be preferable for initial data exploration.

\subsection{Neural network limitations}
Artificial neural networks are powerful non-linear functions that can produce impressive results. Unfortunately, they require the choice of a number of hyperparameters, such as the dimension of the hidden layers and the learning rate, making them less end-user friendly than their classical counterparts.

\section{Conclusion}

We have introduced a new way of capturing the hierarchical properties of scRNA-seq data of a
developing cell population with a density based minimum spanning tree. This tree is a hierarchical representation of the data that places edges in high density regions and thus captures biologically plausible trajectories. The density tree can be used to inform any dimension reduction method about the hierarchical nature of the data.

Moreover, we used the density tree to bias an autoencoder and were thus able to produce promising visualizations exhibiting clearly visible tree-structure both on synthetic and real world scRNA-seq data of developing cell populations.
\section*{Funding}

Supported, in part, by Informatics for Life funded by the Klaus Tschira Foundation.\\

\bibliographystyle{natbib}

\begin{thebibliography}{}

\end{thebibliography}


\begin{thebibliography}{}
\bibitem[Amodio {\em et~al.}(2019)Amodio, van Dijk, Srinivasan, Chen, Mohsen,
	Moon, Campbell, Zhao, Wang, Venkataswamy, Desai, Ravi, Kumar, Montgomery,
	Wolf, and Krishnaswamy]{amodio_exploring_2019}
Amodio, M. et~al (2019).
\newblock Exploring single-cell data with deep multitasking neural networks.
\newblock {\em Nature Methods\/}, {\bf 16}(11), 1139--1145.

\bibitem[Bastidas-Ponce {\em et~al.}(2019)Bastidas-Ponce, Tritschler, Dony,
	Scheibner, Tarquis-Medina, Salinno, Schirge, Burtscher, Böttcher, Theis,
	Lickert, and Bakhti]{bastidas-ponce_comprehensive_2019}
Bastidas-Ponce, A. et~al (2019).
\newblock Comprehensive single cell {mRNA} profiling reveals a detailed roadmap
	for pancreatic endocrinogenesis.
\newblock {\em Development\/}, {\bf 146}(12), dev173849.

\bibitem[Becht {\em et~al.}(2019)Becht, McInnes, Healy, Dutertre, Kwok, Ng,
	Ginhoux, and Newell]{becht_dimensionality_2019}
Becht, E. et~al (2019).
\newblock Dimensionality reduction for visualizing single-cell data using
	{UMAP}.
\newblock {\em Nature Biotechnology\/}, {\bf 37}(1), 38--44.
\newblock Number: 1 Publisher: Nature Publishing Group.

\bibitem[Bendall {\em et~al.}(2011)Bendall, Simonds, Qiu, El-ad, Krutzik,
	Finck, Bruggner, Melamed, Trejo, Ornatsky, {\em et~al.}]{bendall2011single}
Bendall, S.C. et~al (2011).
\newblock Single-cell mass cytometry of differential immune and drug responses
	across a human hematopoietic continuum.
\newblock {\em Science\/}, {\bf 332}(6030), 687--696.

\bibitem[B{\"o}hm {\em et~al.}(2020)B{\"o}hm, Berens, and
	Kobak]{bohm2020attraction}
B{\"o}hm, J.N. et~al (2020).
\newblock Attraction-repulsion spectrum in neighbor embeddings.
\newblock {\em arXiv preprint arXiv:2007.08902\/}.

\bibitem[Cannoodt {\em et~al.}(2016)Cannoodt, Saelens, Sichien, Tavernier,
	Janssens, Guilliams, Lambrecht, Preter, and Saeys]{cannoodt_scorpius_2016}
Cannoodt, R. et~al (2016).
\newblock {SCORPIUS} improves trajectory inference and identifies novel modules
	in dendritic cell development.
\newblock preprint, Bioinformatics.

\bibitem[Cerletti {\em et~al.}(2020)Cerletti, Sandu, Gupta, Oxenius, and
	Claassen]{cerletti_fate_2020}
Cerletti, D. et~al (2020).
\newblock Fate trajectories of {CD8} $^{\textrm{+}}$ {T} cells in chronic
	{LCMV} infection.
\newblock preprint, Immunology.

\bibitem[Damrich and Hamprecht(2021)Damrich and Hamprecht]{damrich_umaps_2021}
Damrich, S. and Hamprecht, F.A. (2021).
\newblock On {UMAP}'s true loss function.
\newblock {\em arXiv:2103.14608 [cs, stat]\/}.
\newblock arXiv: 2103.14608.

\bibitem[Ding {\em et~al.}(2018)Ding, Condon, and Shah]{ding2018interpretable}
Ding, J. et~al (2018).
\newblock Interpretable dimensionality reduction of single cell transcriptome
	data with deep generative models.
\newblock {\em Nature communications\/}, {\bf 9}(1), 1--13.

\bibitem[Eraslan {\em et~al.}(2019)Eraslan, Simon, Mircea, Mueller, and
	Theis]{eraslan2019single}
Eraslan, G. et~al (2019).
\newblock Single-cell rna-seq denoising using a deep count autoencoder.
\newblock {\em Nature communications\/}, {\bf 10}(1), 1--14.

\bibitem[Gr{\o}nbech {\em et~al.}(2020)Gr{\o}nbech, Vording, Timshel,
	S{\o}nderby, Pers, and Winther]{gronbech2020scvae}
Gr{\o}nbech, C.H. et~al (2020).
\newblock scvae: Variational auto-encoders for single-cell gene expression
	data.
\newblock {\em Bioinformatics\/}, {\bf 36}(16), 4415--4422.

\bibitem[Hochgerner {\em et~al.}(2018)Hochgerner, Zeisel, Lönnerberg, and
	Linnarsson]{hochgerner_conserved_2018}
Hochgerner, H. et~al (2018).
\newblock Conserved properties of dentate gyrus neurogenesis across postnatal
	development revealed by single-cell {RNA} sequencing.
\newblock {\em Nature Neuroscience\/}, {\bf 21}(2), 290--299.

\bibitem[Jacomy {\em et~al.}(2014)Jacomy, Venturini, Heymann, and
	Bastian]{jacomy2014forceatlas2}
Jacomy, M. et~al (2014).
\newblock Forceatlas2, a continuous graph layout algorithm for handy network
	visualization designed for the gephi software.
\newblock {\em PloS one\/}, {\bf 9}(6), e98679.

\bibitem[Kingma and Ba(2017)Kingma and Ba]{kingma_adam_2017}
Kingma, D.P. and Ba, J. (2017).
\newblock Adam: A method for stochastic optimization.

\bibitem[Kingma and Welling(2013)Kingma and Welling]{kingma2013auto}
Kingma, D.P. and Welling, M. (2013).
\newblock Auto-encoding variational bayes.
\newblock {\em arXiv preprint arXiv:1312.6114\/}.

\bibitem[Kobak and Berens(2019)Kobak and Berens]{kobak_art_2019}
Kobak, D. and Berens, P. (2019).
\newblock The art of using t-{SNE} for single-cell transcriptomics.
\newblock {\em Nature Communications\/}, {\bf 10}(1), 5416.

\bibitem[Lin {\em et~al.}(2020)Lin, Mukherjee, and Kannan]{lin2020deep}
Lin, E. et~al (2020).
\newblock A deep adversarial variational autoencoder model for dimensionality
	reduction in single-cell rna sequencing analysis.
\newblock {\em BMC bioinformatics\/}, {\bf 21}(1), 1--11.

\bibitem[Lopez {\em et~al.}(2018)Lopez, Regier, Cole, Jordan, and
	Yosef]{lopez2018deep}
Lopez, R. et~al (2018).
\newblock Deep generative modeling for single-cell transcriptomics.
\newblock {\em Nature methods\/}, {\bf 15}(12), 1053--1058.

\bibitem[Luo {\em et~al.}(2021)Luo, Xu, Zhang, and Jin]{luo2021scgae}
Luo, Z. et~al (2021).
\newblock scgae: topology-preserving dimensionality reduction for single-cell
	rna-seq data using graph autoencoder.
\newblock {\em bioRxiv\/}.

\bibitem[Maaten and Hinton(2008)Maaten and Hinton]{maaten_visualizing_2008}
Maaten, L.v.d. and Hinton, G. (2008).
\newblock Visualizing {Data} using t-{SNE}.
\newblock {\em Journal of Machine Learning Research\/}, {\bf 9}(86),
	2579--2605.

\bibitem[Martinetz and Schulten(1994)Martinetz and
	Schulten]{martinetz_topology_1994}
Martinetz, T. and Schulten, K. (1994).
\newblock Topology representing networks.
\newblock {\em Neural Networks\/}, {\bf 7}(3), 507--522.

\bibitem[McInnes {\em et~al.}(2020)McInnes, Healy, and
	Melville]{mcinnes_umap_2020}
McInnes, L. et~al (2020).
\newblock {UMAP}: {Uniform} {Manifold} {Approximation} and {Projection} for
	{Dimension} {Reduction}.
\newblock {\em arXiv:1802.03426 [cs, stat]\/}.
\newblock arXiv: 1802.03426.

\bibitem[Moon {\em et~al.}(2019)Moon, van Dijk, Wang, Gigante, Burkhardt, Chen,
	Yim, Elzen, Hirn, Coifman, Ivanova, Wolf, and
	Krishnaswamy]{moon_visualizing_2019}
Moon, K.R. et~al (2019).
\newblock Visualizing structure and transitions in high-dimensional biological
	data.
\newblock {\em Nature Biotechnology\/}, {\bf 37}(12), 1482--1492.

\bibitem[Moor {\em et~al.}(2020)Moor, Horn, Rieck, and
	Borgwardt]{moor_topological_2020}
Moor, M. et~al (2020).
\newblock Topological {Autoencoders}.
\newblock {\em arXiv:1906.00722 [cs, math, stat]\/}.
\newblock arXiv: 1906.00722.

\bibitem[Paszke {\em et~al.}(2019)Paszke, Gross, Massa, Lerer, Bradbury,
	Chanan, Killeen, Lin, Gimelshein, Antiga, Desmaison, Köpf, Yang, DeVito,
	Raison, Tejani, Chilamkurthy, Steiner, Fang, Bai, and
	Chintala]{paszke_pytorch_2019}
Paszke, A. et~al (2019).
\newblock {PyTorch}: {An} {Imperative} {Style}, {High}-{Performance} {Deep}
	{Learning} {Library}.
\newblock {\em arXiv:1912.01703 [cs, stat]\/}.
\newblock arXiv: 1912.01703.

\bibitem[Perret {\em et~al.}(2019)Perret, Chierchia, Cousty, F. Guimarães,
	Kenmochi, and Najman]{perret_higra_2019}
Perret, B. et~al (2019).
\newblock Higra: {Hierarchical} {Graph} {Analysis}.
\newblock {\em SoftwareX\/}, {\bf 10}, 100335.

\bibitem[Qiu {\em et~al.}(2011)Qiu, Simonds, Bendall, Gibbs, Bruggner,
	Linderman, Sachs, Nolan, and Plevritis]{qiu2011extracting}
Qiu, P. et~al (2011).
\newblock Extracting a cellular hierarchy from high-dimensional cytometry data
	with spade.
\newblock {\em Nature biotechnology\/}, {\bf 29}(10), 886--891.

\bibitem[Qiu {\em et~al.}(2017)Qiu, Mao, Tang, Wang, Chawla, Pliner, and
	Trapnell]{qiu_reversed_2017}
Qiu, X. et~al (2017).
\newblock Reversed graph embedding resolves complex single-cell trajectories.
\newblock {\em Nature Methods\/}, {\bf 14}(10), 979--982.

\bibitem[Rumelhart {\em et~al.}(1985)Rumelhart, Hinton, and
	Williams]{rumelhart_learning_1985}
Rumelhart, D.E. et~al (1985).
\newblock Learning internal representations by error propagation.
\newblock Technical report, California Univ San Diego La Jolla Inst for
	Cognitive Science.

\bibitem[Saelens {\em et~al.}(2019)Saelens, Cannoodt, Todorov, and
	Saeys]{saelens_comparison_2019}
Saelens, W. et~al (2019).
\newblock A comparison of single-cell trajectory inference methods.
\newblock {\em Nature Biotechnology\/}, {\bf 37}(5), 547--554.

\bibitem[Street {\em et~al.}(2018)Street, Risso, Fletcher, Das, Ngai, Yosef,
	Purdom, and Dudoit]{street_slingshot_2018}
Street, K. et~al (2018).
\newblock Slingshot: cell lineage and pseudotime inference for single-cell
	transcriptomics.
\newblock {\em BMC Genomics\/}, {\bf 19}(1), 477.

\bibitem[Szubert {\em et~al.}(2019)Szubert, Cole, Monaco, and
	Drozdov]{szubert2019structure}
Szubert, B. et~al (2019).
\newblock Structure-preserving visualisation of high dimensional single-cell
	datasets.
\newblock {\em Scientific reports\/}, {\bf 9}(1), 1--10.

\bibitem[Tian {\em et~al.}(2019)Tian, Wan, Song, and Wei]{tian2019clustering}
Tian, T. et~al (2019).
\newblock Clustering single-cell rna-seq data with a model-based deep learning
	approach.
\newblock {\em Nature Machine Intelligence\/}, {\bf 1}(4), 191--198.

\bibitem[Waddington(1957)Waddington]{waddington_strategy_1957}
Waddington, C.H. (1957).
\newblock {\em The strategy of the genes : a discussion of some aspects of
	theoretical biology\/}.
\newblock Routledge {Library} {Editions}: 20th {Century} {Science}. Routledge.

\bibitem[Wolf {\em et~al.}(2019)Wolf, Hamey, Plass, Solana, Dahlin, Göttgens,
	Rajewsky, Simon, and Theis]{wolf_paga_2019}
Wolf, F.A. et~al (2019).
\newblock {PAGA}: graph abstraction reconciles clustering with trajectory
	inference through a topology preserving map of single cells.
\newblock {\em Genome Biology\/}, {\bf 20}(1), 59.

\bibitem[Zheng {\em et~al.}(2017)Zheng, Terry, Belgrader, Ryvkin, Bent, Wilson,
	Ziraldo, Wheeler, McDermott, Zhu, Gregory, Shuga, Montesclaros, Underwood,
	Masquelier, Nishimura, Schnall-Levin, Wyatt, Hindson, Bharadwaj, Wong, Ness,
	Beppu, Deeg, McFarland, Loeb, Valente, Ericson, Stevens, Radich, Mikkelsen,
	Hindson, and Bielas]{zheng_massively_2017}
Zheng, G.X.Y. et~al (2017).
\newblock Massively parallel digital transcriptional profiling of single cells.
\newblock {\em Nature Communications\/}, {\bf 8}(1).

\end{thebibliography}

\clearpage
\appendix
\setcounter{figure}{0}
\setcounter{table}{0}
\setcounter{algorithm}{0}
 \renewcommand{\thefigure}{S\arabic{figure}}
  \renewcommand{\thetable}{S\arabic{table}}
    \renewcommand{\thealgorithm}{S\arabic{algorithm}}
\section{Training loop algorithm}
\begin{algorithm}[h]
	\caption{Training loop}
	\begin{algorithmic}[1]
		\Require  Autoencoder $(g \circ f)_\theta$
		\Require  Pretraining epochs $n_p$, batch size $b$ and learning rate $\alpha_p$
		\Require  Finetuning epochs $n_f$ and learning rate $\alpha_f$
		\Require  Weight parameters for the loss
		$\lambda_{\text{rec}}$,$\lambda_{\text{push-pull}}$,$\lambda_{\text{comp}}$,$\lambda_{\text{cos}}$
		\State $ T, C , C_2, d_{geo}\gets \Call{Initialization}{X}$
		\State $\# Pretraining$
		\For{$t = 0,1,\ldots,n_p$}
			\For{$i = 0,1,\ldots,n_p/b$}
				\State Sample a minibatch $m$ from $X$
				\State $\hat{m} \gets g(f(m))$
				\State $\mathcal{L} \gets \mathcal{L}_{\text{rec}}$
				\State $ \theta^{t+1} \gets \theta^t - \alpha_p \nabla \mathcal{L}$
			\EndFor
		\EndFor
		\State $\# Finetuning$
		\For{$t = n_p,\ldots,n_p+n_f$}
			\State $h \gets f(X)$
			\State $\hat{X} \gets g(h)$
			\State $\mathcal{L} \gets \lambda_{\text{rec}}\mathcal{L}_{\text{rec}} + \lambda_{\text{push-pull}}\mathcal{L}_{\text{push-pull}} +
	\lambda_{\text{comp}}\mathcal{L}_{\text{comp}} + \lambda_{\text{cos}}\mathcal{L}_{\text{cos}}$
			\State $ \theta^{t+1} \gets \theta^t - \alpha_f \nabla \mathcal{L}$
		\EndFor
	\end{algorithmic}
	\label{algo:training}
\end{algorithm}
\section{Cosine loss generalization}

The definition of a vertex' degree in a graph as the number of incident edges to it is not perfect, as it does not take into account the noisiness of the graph. On real datasets, we may have stray clusters which lead to noisy edges in the density graph. These usually manifest as edges with only one point contributing to them in high dimension. This leads to vertices with an effective degree of 2 that have a higher degree due to these noisy edges, and are thus ignored by the cosine loss.

To remedy this, we introduce a different definition of degree.
We consider a threshold $t \in [0,100]$ and define the degree of a vertex as the smallest number of incident edges that account for $t\%$ of all points contributing to the vertex's incident edges. As $t$ gets closer to a hundred, we converge to the original definition of degree. 

More formally put, consider a weighted graph \mbox{$G = (V,E,W)$} and a function $\Gamma$ that returns incident edges to a given vertex sorted by their weights.
This alternative definition of a vertex's degree is then:

\begin{align*}
    \text{deg}(v,t) = &\min_{n=1\ldots|\Gamma(v)|} \quad n \\
	&\text{s.t.}\quad \frac{\sum_{i=1}^{n} W_{\Gamma(v)_i}}{\sum_{j=1}^{|\Gamma(v)|} W_{\Gamma(v)_j}} \geq \frac{t}{100}
\end{align*}

We can clearly see that when $t=100$ we obtain the classical definition of degree.
As this generalization has not improved the visualization quality drastically, we opted for the simpler version of the cosine loss in the main paper.

\section{Ablation study}\label{sec:ablation}

In order to better visualize the contributions of each element of our method, we conducted an ablation study of the different loss parameters and evaluated their impact both qualitatively and quantitatively.

\subsection{Loss parameters}

The first phenomenon that is studied is the influence of dropping loss terms entirely. The reconstruction loss is always kept since it is necessary for the embeddings to contain salient information about the data. Not all combinations of loss parameters will be studied, but only those that should be interesting (for example, using only the cosine loss does not make much sense, so it is not an interesting scenario).

We will not study the influence of the weights for every loss since the default weights of $1$ lead to good performance and this configuration significantly reduces the dimension of the hyperparameter space. All experiments are described in table~\ref{tab:ablation}.

The performance will be evaluated both qualitatively and quantitatively on all three discussed datasets to demonstrate as clearly as possible the impact of every loss term.

\begin{table}[h!]
\centering
 \begin{tabular}{l c c c c} 
 \toprule
 Experiment & $\mathcal{L}_{\text{rec}}$ & $\mathcal{L}_{\text{push-pull}}$ & $\mathcal{L}_{\text{comp}}$ & $\mathcal{L}_{\text{cos}}$ (weight)\\
 \hline
 A & \checkmark & & & \\
 B & \checkmark & & \checkmark & \\
 C & \checkmark & \checkmark & & \\
 D & \checkmark & \checkmark & \checkmark & \\
 E & \checkmark & \checkmark & & \checkmark (50) \\
 F & \checkmark & \checkmark &\checkmark & \checkmark (50)\\
 \bottomrule
 \end{tabular}
 \caption{List of loss parameters for our ablations.}
 \label{tab:ablation}
\end{table}

\begin{figure}[t!]
	\centering 
	\includegraphics[width=0.9\linewidth]{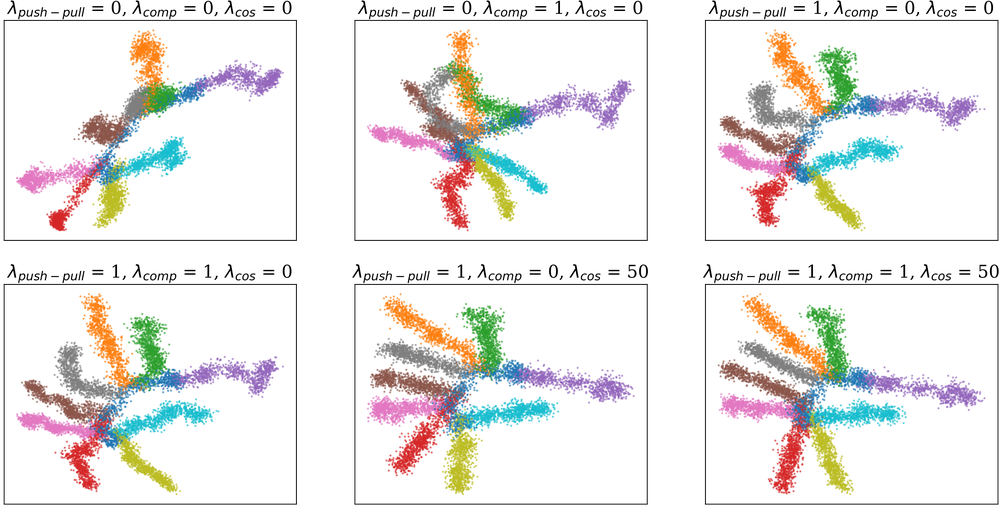}
	\caption{Results of the ablations on the PHATE generated dataset, colored by groundtruth clusters.}%
	\label{fig:ablation-phate} 
\end{figure}

\begin{figure}[t!]
	\centering 
	\includegraphics[width=0.9\linewidth]{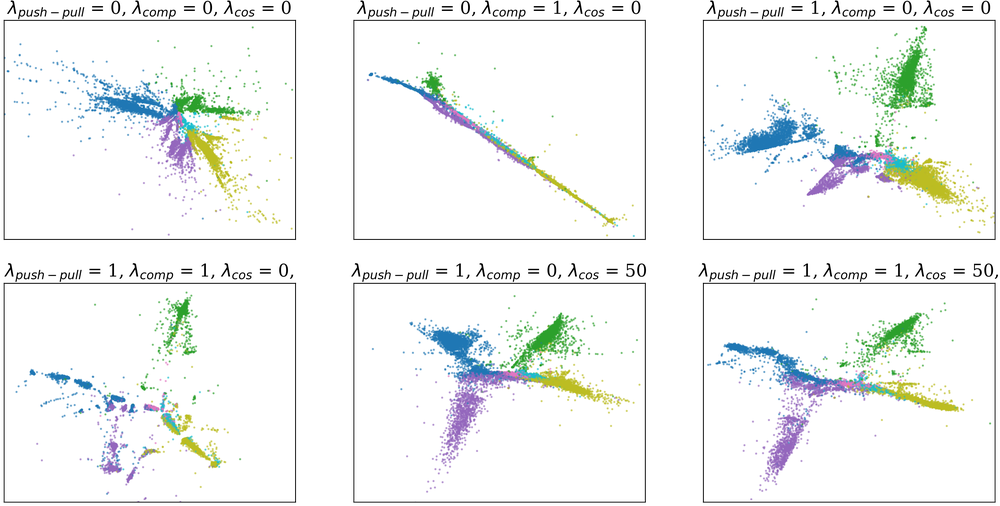}
	\caption{Results of the ablations on the T-cell dataset, colored by phenotypes.}%
	\label{fig:ablation-t-cells} 
\end{figure}

\begin{figure}[t!]
	\centering 
	\includegraphics[width=0.9\linewidth]{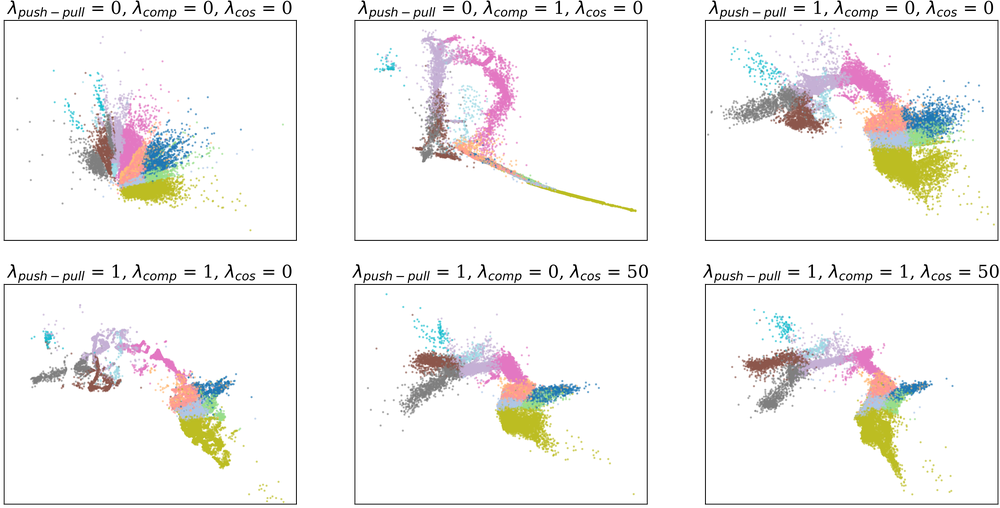}
	\caption{Results of the ablations on the endocrine pancreas dataset, colored by cell types.}%
	\label{fig:ablation-pancreatic} 
\end{figure}

\begin{table*}[t!]

        \begin{subtable}{\textwidth}
        \centering
         \begin{tabular}{l c c c c c c c c } 
         \toprule
          Type of metric & \multicolumn{2}{c}{Local} & \multicolumn{4}{c}{Global} & \multicolumn{2}{c}{Voronoi}\\
           \cmidrule(l{2pt}r{2pt}){2-3} \cmidrule(l{2pt}r{2pt}){4-7} \cmidrule(l{2pt}r{2pt}){8-9} 
          \multirow{2}{*}{Metric}&\multirow{2}{*}{ARI}&\multirow{2}{*}{k-NN}& \multicolumn{2}{c}{Euclidean} & \multicolumn{2}{c}{Geodesic} &\multirow{2}{*}{1\textsuperscript{st} order} & \multirow{2}{*}{2\textsuperscript{nd} order}  \\
         \cmidrule(l{2pt}r{2pt}){4-5} \cmidrule(l{2pt}r{2pt}){6-7} 
           &  & & Pearson & Spearman & Pearson & Spearman &   &  \\
         \hline
        $\lambda_{pp} = 0, \lambda_{comp} = 0, \lambda_{cos} = 0$ & 34.62  & 19.09  & 0.66  & 0.66  & 0.56  & 0.56  & 70.01  & 38.98 \\ 
$\lambda_{pp} = 0, \lambda_{comp} = 1, \lambda_{cos} = 0$ & 38.54  & 23.20  & 0.80  & 0.79  & 0.74  & 0.73  & 70.44  & 35.24 \\ 
$\lambda_{pp} = 1, \lambda_{comp} = 0, \lambda_{cos} = 0$ & 48.19  & 25.05  & 0.78  & 0.76  & 0.72  & 0.70  & \textbf{79.68}  & \textbf{56.02} \\ 
$\lambda_{pp} = 1, \lambda_{comp} = 1, \lambda_{cos} = 0$ & \textbf{48.70}  & \textbf{26.40}  & \textbf{0.81}  & 0.78  & 0.74  & 0.72  & 79.26  & 55.93 \\ 
$\lambda_{pp} = 1, \lambda_{comp} = 0, \lambda_{cos} = 50$ & 45.54  & 22.71  & 0.80  & 0.77  & 0.74  & 0.72  & 78.94  & 52.97 \\ 
$\lambda_{pp} = 1, \lambda_{comp} = 1, \lambda_{cos} = 50$ & 46.00  & 24.60  & \textbf{0.81}  & \textbf{0.80}  & \textbf{0.75}  & \textbf{0.74}  & 78.85  & 53.77 \\ 

     \bottomrule
     \end{tabular}
     
\caption{PHATE generated dataset.}
\label{quant_phate}
     \end{subtable}
     
\begin{subtable}{\textwidth}
        \centering
         \begin{tabular}{l c c c c c c c c } 
        \toprule
         \multirow{2}{*}{Metric}&\multirow{2}{*}{ARI}&\multirow{2}{*}{k-NN}& \multicolumn{2}{c}{Euclidean} & \multicolumn{2}{c}{Geodesic} &\multirow{2}{*}{1\textsuperscript{st} order} & \multirow{2}{*}{2\textsuperscript{nd} order}  \\
        \cmidrule(l{2pt}r{2pt}){4-5} \cmidrule(l{2pt}r{2pt}){6-7} &  & & Pearson & Spearman & Pearson & Spearman &   &  \\
         \hline
        $\lambda_{pp} = 0, \lambda_{comp} = 0, \lambda_{cos} = 0$ & 34.52  & \textbf{4.17}  & \textbf{0.81}  & \textbf{0.85}  & 0.57  & 0.62  & 65.37  & 30.83 \\ 
$\lambda_{pp} = 0, \lambda_{comp} = 1, \lambda_{cos} = 0$ & 24.88  & 2.32  & 0.65  & 0.69  & 0.53  & 0.59  & 46.30  & 11.11 \\ 
$\lambda_{pp} = 1, \lambda_{comp} = 0, \lambda_{cos} = 0$ & 43.07  & 3.07  & 0.77  & 0.79  & \textbf{0.71}  & 0.77  & \textbf{73.79}  & \textbf{50.68} \\ 
$\lambda_{pp} = 1, \lambda_{comp} = 1, \lambda_{cos} = 0$ & \textbf{44.69}  & 2.93  & 0.73  & 0.79  & 0.66  & 0.75  & 72.95  & 46.75 \\ 
$\lambda_{pp} = 1, \lambda_{comp} = 0, \lambda_{cos} = 50$ & 35.64  & 2.77  & 0.73  & 0.75  & 0.70  & 0.75  & 68.44  & 38.82 \\ 
$\lambda_{pp} = 1, \lambda_{comp} = 1, \lambda_{cos} = 50$ & 39.79  & 2.85  & 0.71  & 0.74  & \textbf{0.71}  & \textbf{0.78}  & 69.24  & 38.04 \\ 

     \bottomrule
     \end{tabular}
     
\caption{Endocrine pancreas dataset.}
\label{quant_pancreas}
     \end{subtable}

        \begin{subtable}{\textwidth}
        \centering
         \begin{tabular}{l c c c c c c c c } 
        \toprule
         \multirow{2}{*}{Metric}&\multirow{2}{*}{ARI}&\multirow{2}{*}{k-NN}& \multicolumn{2}{c}{Euclidean} & \multicolumn{2}{c}{Geodesic} &\multirow{2}{*}{1\textsuperscript{st} order} & \multirow{2}{*}{2\textsuperscript{nd} order}  \\
        \cmidrule(l{2pt}r{2pt}){4-5} \cmidrule(l{2pt}r{2pt}){6-7} &  & & Pearson & Spearman & Pearson & Spearman &   &  \\
         \hline
        $\lambda_{pp} = 0, \lambda_{comp} = 0, \lambda_{cos} = 0$ & 29.24  & \textbf{2.20}  & 0.40  & \textbf{0.33}  & 0.40  & 0.42  & 35.17  & 4.50 \\ 
$\lambda_{pp} = 0, \lambda_{comp} = 1, \lambda_{cos} = 0$ & 40.65  & 1.24  & 0.15  & 0.17  & 0.20  & 0.20  & 28.75  & 2.75 \\ 
$\lambda_{pp} = 1, \lambda_{comp} = 0, \lambda_{cos} = 0$ & 29.72  & 1.28  & \textbf{0.42}  & 0.26  & 0.36  & 0.39  & \textbf{47.19}  & \textbf{18.63} \\ 
$\lambda_{pp} = 1, \lambda_{comp} = 1, \lambda_{cos} = 0$ & \textbf{45.55}  & 1.15  & 0.38  & 0.23  & 0.35  & 0.38  & 37.71  & 16.29 \\ 
$\lambda_{pp} = 1, \lambda_{comp} = 0, \lambda_{cos} = 50$ & 29.24  & 1.23  & 0.37  & 0.24  & \textbf{0.42}  & \textbf{0.44}  & 44.73  & 12.16 \\ 
$\lambda_{pp} = 1, \lambda_{comp} = 1, \lambda_{cos} = 50$ & 37.25  & 1.15  & 0.31  & 0.19  & 0.40  & 0.41  & 38.41  & 12.85 \\ 

     \bottomrule
     \end{tabular}
     
\caption{T-cells dataset.}
\label{quant_t-cells}
     \end{subtable}

 \caption{Quantitative results in different scenarios for DTAE's loss weights.}
 \label{tab:ablation_quant}
\end{table*}

\begin{table}[h!]
\centering
 \begin{tabular}{l c} 
 \toprule
   & Rel. Perf. \\
 \hline
 $\lambda_{pp} = 0, \lambda_{comp} = 0, \lambda_{cos} = 0$ & 81.27 \\ 
$\lambda_{pp} = 0, \lambda_{comp} = 1, \lambda_{cos} = 0$ & 67.60 \\ 
$\lambda_{pp} = 1, \lambda_{comp} = 0, \lambda_{cos} = 0$ & \textbf{92.04} \\ 
$\lambda_{pp} = 1, \lambda_{comp} = 1, \lambda_{cos} = 0$ & 90.66 \\ 
$\lambda_{pp} = 1, \lambda_{comp} = 0, \lambda_{cos} = 50$ & 87.24 \\ 
$\lambda_{pp} = 1, \lambda_{comp} = 1, \lambda_{cos} = 50$ & 86.99 \\ 

 \bottomrule
 \end{tabular}
 \caption{Relative performance out of a hundred over all datasets and metrics.}
 \label{tab:ablation_quant_aggregated}
\end{table}

\begin{table*}[t!]
        \centering
         \begin{tabular}{l c c c c c c c c c} 
         \toprule
          Type of metric & \multicolumn{2}{c}{Local} & \multicolumn{4}{c}{Global} & \multicolumn{2}{c}{Voronoi} & \\
           \cmidrule(l{2pt}r{2pt}){2-3} \cmidrule(l{2pt}r{2pt}){4-7} \cmidrule(l{2pt}r{2pt}){8-9} 
          \multirow{2}{*}{Metric}&\multirow{2}{*}{ARI}&\multirow{2}{*}{k-NN}& \multicolumn{2}{c}{Euclidean} & \multicolumn{2}{c}{Geodesic} &\multirow{2}{*}{1\textsuperscript{st} order} & \multirow{2}{*}{2\textsuperscript{nd} order} & All \\
         \cmidrule(l{2pt}r{2pt}){4-5} \cmidrule(l{2pt}r{2pt}){6-7} 
           &  & & Pearson & Spearman & Pearson & Spearman &   &  \\
         \hline
        $\lambda_{cos} = 1$ & \textbf{45.83}  & 1.15  & \textbf{0.37}  & \textbf{0.24}  & 0.38  & 0.39  & 37.30  & 16.36 & \textbf{95.68}\\ 
$\lambda_{cos} = 2$ & 44.77  & 1.11  & 0.35  & \textbf{0.24}  & 0.40  & 0.40  & 37.90  & \textbf{16.39} & 95.34 \\ 
$\lambda_{cos} = 5$ & 45.22  & 1.09  & \textbf{0.37}  & 0.20  & 0.40  & 0.45  & 37.38  & 14.12 & 93.30 \\ 
$\lambda_{cos} = 10$ & 44.29  & 1.12  & 0.35  & 0.18  & 0.42  & \textbf{0.46}  & 36.40  & 13.66 & 91.83\\ 
$\lambda_{cos} = 15$ & 43.50  & 1.14  & 0.29  & 0.15  & \textbf{0.44}  & \textbf{0.46}  & \textbf{38.78}  & 14.59 & 90.28 \\ 
$\lambda_{cos} = 20$ & 43.08  & \textbf{1.17}  & 0.33  & 0.17  & 0.42  & \textbf{0.46}  & 37.43  & 14.41 & 91.74 \\ 
$\lambda_{cos} = 50$ & 37.25  & 1.15  & 0.31  & 0.19  & 0.40  & 0.41  & 38.41  & 12.85 & 87.50 \\ 

     \bottomrule
     \end{tabular}

 \caption{Quantitative results on the T-cells dataset when varying the cosine loss weight. The weights for the push-pull and compactness losses are set to one. The rightmost column contains the average performance over all metrics for a given method.}
 \label{tab:cosine_quant}
\end{table*}

As can be seen in figures~\ref{fig:ablation-phate},\ref{fig:ablation-t-cells} and~\ref{fig:ablation-pancreatic}, the compactness loss alone is not sufficient to obtain a good representation since it has no repulsive force. The reconstruction loss helps to avoid a total collapse but is not sufficient to prevent a partial collapse, as visible in the endocrine pancreas and the T-cell datasets.
While the push-pull loss already gives good results when used alone, since the tree structure is visible, adding the compactness loss yields embeddings in which the points lie compactly along the tree. Without the cosine loss, however, this combination can lead to sparse representation due to the fact that seeds of second order Voronoï cells do not necessarily lie in their cell. This means that points will not necessarily be spread out along the line between two centroids but only lie inside the intersection of the line between the two centroids and their second order Voronoï cell, which may be much smaller than the full line between the centroids.
Using only the push-pull and cosine loss can lead to satisfying results, but the embedding is more spread out than with the compactness loss.
Adding the cosine loss makes all the results cleaner and helps with the density of the point cloud. This effect is discussed in the next section.

From a quantitative point of view, adding all of these losses leads to worse performances than just using the push-pull loss alone. Since the compactness and cosine losses are designed with visualization in mind, they can alter the fidelity of the embedding. For example, making the points tighter along the density tree will lead to pairwise distances that are preserved more poorly, which is an effect that we indeed observe in the global metrics in table~\ref{tab:ablation_quant}.\\
Nonetheless, when looking at aggregated performances in table~\ref{tab:ablation_quant_aggregated} we can see that all experiments except when using the compactness loss alone still perform comparatively. As such, the increase in qualitative performance stemming from the addition of losses is not done at the expense of the preservation of the data's intrinsic structure. In particular, the push-pull loss alone drastically improves the visualization not only qualitatively, but also quantitatively.

\subsection{Cosine loss weight}

A parameter that is interesting to study in more detail is the cosine loss weight. While a lot of the other losses have a significant impact on the embeddings, the cosine loss is mostly cosmetic, and it is important to understand its behavior for low and high weights.
The cosine loss weight will only be studied on the T-cell dataset, since it is enough to demonstrate its impact on quantitative and qualitative results.

\begin{figure}[t!]
	\centering 
	\includegraphics[width=1\linewidth]{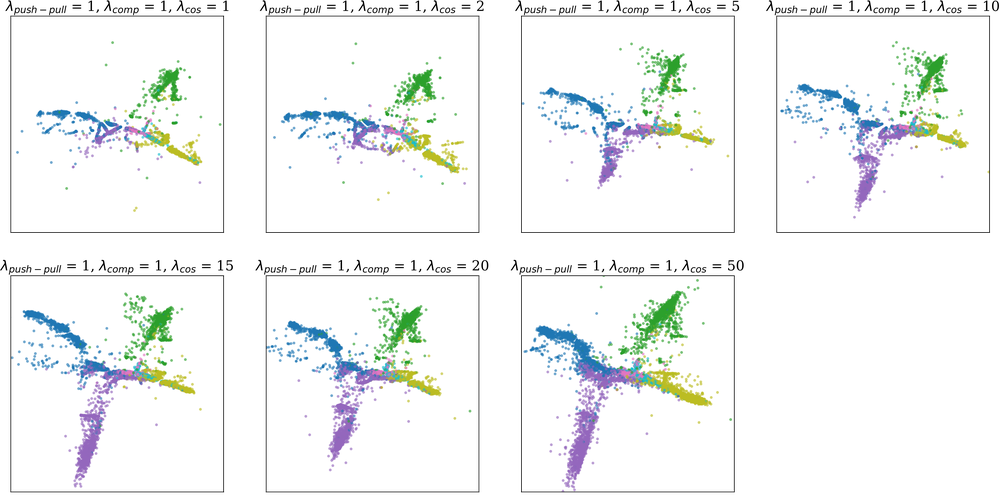}
	\caption{Results obtained on the chronic infection subset of the T-cell dataset when varying the cosine loss weight, colored by phenotypes.}%
	\label{fig:ablation-cosine} 
\end{figure}

As can be seen in figure~\ref{fig:ablation-cosine} the cosine loss straightens the branches for every weight as intended. However, with higher weights, it also has a density regularizing effect. As its weight increases, we obtain a more homogeneous and less clumped point cloud.
While there is no clear explanation for this behavior, a hypothesis is that the higher weight means that this criterion will be optimized with higher priority during the finetuning. Since the pretraining produces dense embedding and this cosine loss has no incentive to produce sparse embeddings, this denser structure is kept during training.
On the contrary, the push-pull loss can have a sparsifying effect, since the seeds of second order Voronoï cells do not necessarily lie in their cells. When the cosine loss weight is smaller, this loss is optimized with higher priority, which would lead to the sparser embeddings.
All of this is intimately linked to the dynamics of neural network training and not only to minimizers of each criterion, making a precise study of this process highly complex.

From a quantitative point of view, a slight decrease in performance is visible in table~\ref{tab:cosine_quant} for all metrics except for the preservation of geodesic distances or of first order Voronoï diagrams. As a result, the overall performance decreases noticeably when increasing the cosine loss weight, see the rightmost column in table~\ref{tab:cosine_quant}.

This again illustrates the trade-offs between quantitative and qualitative performance, where even though a method performs slightly worse quantitatively, it might still produce results that are easier to interpret for humans.

\clearpage
\onecolumn
\section{High resolution results}
\begin{figure}[h!]
	\centering 
	\includegraphics[width=1\linewidth]{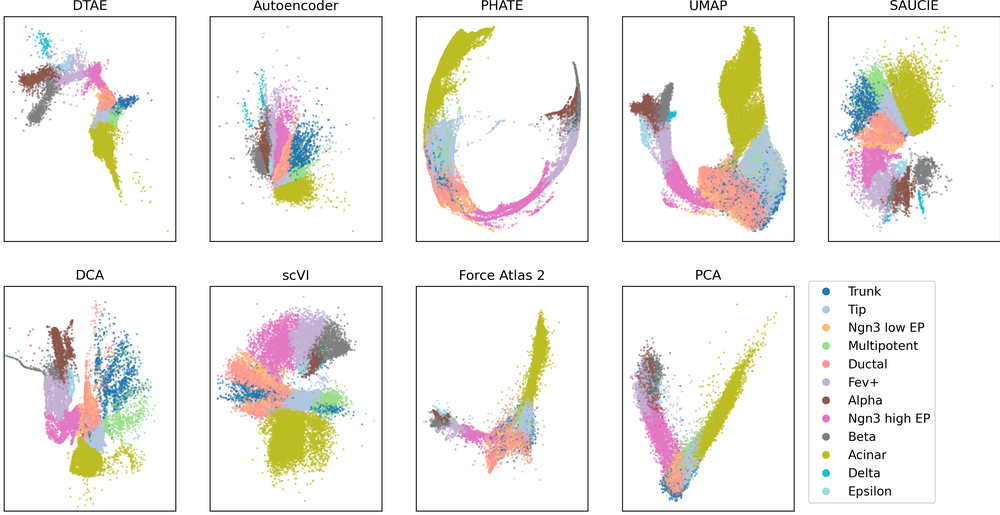}
		\caption{Results obtained on the endocrine pancreatic cell dataset, colored by cell types.}
	\label{fig:results-pancreas} 
\end{figure}

\begin{figure}[h!]
	\centering 
	\includegraphics[width=1\linewidth]{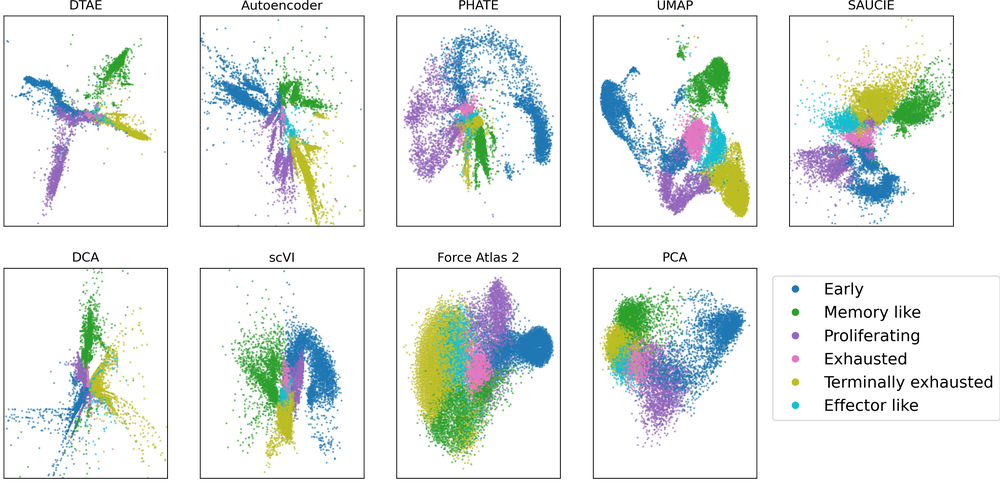}
	\caption{Results obtained on the chronic infection subset of the T-cell dataset, colored by phenotypes.}%
	\label{fig:results-eth-chronic-louvain} 
\end{figure}

\clearpage
\section {Complete quantitative results}
\begin{table}[h!]

        \begin{subtable}{\textwidth}
        \centering
         \begin{tabular}{l c c c c c c c c } 
         \toprule
          Type of metric & \multicolumn{2}{c}{Local} & \multicolumn{4}{c}{Global} & \multicolumn{2}{c}{Voronoi}\\
           \cmidrule(l{2pt}r{2pt}){2-3} \cmidrule(l{2pt}r{2pt}){4-7} \cmidrule(l{2pt}r{2pt}){8-9} 
          \multirow{2}{*}{Metric}&\multirow{2}{*}{ARI}&\multirow{2}{*}{k-NN}& \multicolumn{2}{c}{Euclidean} & \multicolumn{2}{c}{Geodesic} &\multirow{2}{*}{1\textsuperscript{st} order} & \multirow{2}{*}{2\textsuperscript{nd} order}  \\
         \cmidrule(l{2pt}r{2pt}){4-5} \cmidrule(l{2pt}r{2pt}){6-7} 
           &  & & Pearson & Spearman & Pearson & Spearman &   &  \\
         \hline
        DTAE (Ours) & 46.00  & 24.60  & \textbf{0.81}  & \textbf{0.80}  & \textbf{0.75}  & \textbf{0.74}  & 78.85  & 53.77 \\ 
AE & 34.67  & 19.18  & 0.63  & 0.64  & 0.55  & 0.54  & 70.56  & 38.64 \\ 
PHATE & 51.33  & 60.44  & 0.50  & 0.46  & 0.54  & 0.52  & 71.12  & 30.40 \\ 
UMAP & 55.13  & \textbf{67.92}  & 0.53  & 0.48  & 0.54  & 0.51  & 75.18  & 46.19 \\ 
SAUCIE & \textbf{56.62}  & 37.61  & \textbf{0.81}  & 0.79  & \textbf{0.75}  & 0.73  & \textbf{82.98}  & \textbf{65.07} \\ 
DCA & 40.41  & 21.29  & 0.64  & 0.64  & 0.59  & 0.59  & 73.83  & 43.17 \\ 
scVI & 36.51  & 19.68  & 0.69  & 0.68  & 0.67  & 0.67  & 69.94  & 36.98 \\ 
Force Atlas 2 & 51.85  & 64.38  & 0.59  & 0.55  & 0.56  & 0.54  & 75.79  & 46.44 \\ 
PCA & 39.87  & 22.36  & 0.77  & 0.74  & 0.67  & 0.66  & 73.14  & 42.53 \\ 

     \bottomrule
     \end{tabular}
     
\caption{PHATE generated dataset.}
\label{tab:quant_phate}
     \end{subtable}

         \begin{subtable}{\textwidth}
        \centering
         \begin{tabular}{l c c c c c c c c } 
        \toprule
         \multirow{2}{*}{Metric}&\multirow{2}{*}{ARI}&\multirow{2}{*}{k-NN}& \multicolumn{2}{c}{Euclidean} & \multicolumn{2}{c}{Geodesic} &\multirow{2}{*}{1\textsuperscript{st} order} & \multirow{2}{*}{2\textsuperscript{nd} order}  \\
        \cmidrule(l{2pt}r{2pt}){4-5} \cmidrule(l{2pt}r{2pt}){6-7} &  & & Pearson & Spearman & Pearson & Spearman &   &  \\
         \hline
        DTAE (Ours) & \textbf{39.79}  & 2.85  & 0.71  & 0.74  & 0.71  & 0.78  & 69.24  & \textbf{38.04} \\ 
AE & 34.12  & 4.24  & \textbf{0.81}  & \textbf{0.84}  & 0.58  & 0.62  & 65.12  & 30.53 \\ 
PHATE & 30.92  & 3.70  & 0.64  & 0.65  & 0.71  & 0.78  & 57.22  & 22.27 \\ 
UMAP & 30.41  & 4.67  & 0.57  & 0.58  & 0.79  & 0.82  & 57.79  & 21.21 \\ 
SAUCIE & 38.94  & 3.69  & \textbf{0.81}  & 0.81  & 0.71  & 0.73  & \textbf{69.46}  & 37.93 \\ 
DCA & 30.93  & \textbf{5.01}  & 0.41  & 0.78  & 0.37  & 0.63  & 64.29  & 27.98 \\ 
scVI & 32.99  & 3.80  & 0.67  & 0.68  & 0.65  & 0.67  & 61.29  & 27.52 \\ 
Force Atlas 2 & 25.11  & 3.96  & 0.28  & 0.62  & 0.21  & 0.74  & 46.24  & 13.08 \\ 
PCA & 21.84  & 1.94  & 0.69  & 0.67  & \textbf{0.87}  & \textbf{0.87}  & 48.65  & 15.50 \\ 

     \bottomrule
     \end{tabular}
     
\caption{Endocrine pancreas dataset.}
\label{tab:quant_pancreas}
     \end{subtable}    
        \begin{subtable}{\textwidth}
        \centering
         \begin{tabular}{l c c c c c c c c } 
        \toprule
         \multirow{2}{*}{Metric}&\multirow{2}{*}{ARI}&\multirow{2}{*}{k-NN}& \multicolumn{2}{c}{Euclidean} & \multicolumn{2}{c}{Geodesic} &\multirow{2}{*}{1\textsuperscript{st} order} & \multirow{2}{*}{2\textsuperscript{nd} order}  \\
        \cmidrule(l{2pt}r{2pt}){4-5} \cmidrule(l{2pt}r{2pt}){6-7} &  & & Pearson & Spearman & Pearson & Spearman &   &  \\
         \hline
        DTAE (Ours) & \textbf{37.25}  & 1.15  & 0.31  & 0.19  & 0.40  & 0.41  & \textbf{38.41}  & \textbf{12.85} \\ 
AE & 28.87  & \textbf{2.17}  & 0.38  & 0.32  & 0.43  & 0.43  & 34.84  & 4.58 \\ 
PHATE & 32.00  & 1.25  & -0.02  & 0.02  & 0.42  & 0.43  & 33.70  & 3.45 \\ 
UMAP & 23.41  & 1.52  & 0.11  & 0.21  & 0.46  & 0.44  & 29.24  & 5.28 \\ 
SAUCIE & 26.86  & 1.59  & 0.21  & 0.25  & 0.43  & 0.42  & 34.30  & 4.63 \\ 
DCA & 0.10  & 1.34  & \textbf{0.45}  & \textbf{0.62}  & 0.00  & 0.26  & 3.17  & 1.04 \\ 
scVI & 28.69  & 1.26  & 0.43  & 0.23  & 0.38  & 0.46  & 33.31  & 5.67 \\ 
Force Atlas 2 & 23.83  & 0.93  & 0.02  & 0.01  & 0.05  & 0.41  & 28.39  & 3.09 \\ 
PCA & 20.82  & 1.10  & 0.18  & 0.16  & \textbf{0.61}  & \textbf{0.57}  & 32.30  & 8.27 \\ 

     \bottomrule
     \end{tabular}
     
\caption{T-cells dataset.}
\label{tab:quant_t-cells}
     \end{subtable}

 \caption{Full Quantitative results on all studied datasets. Metrics are described in section~\ref{sec:quantitative} and higher values indicate better performance.}
 \label{tab:quantitative-full}
\end{table}

\twocolumn

\end{document}